\documentclass[runningheads]{llncs}

\usepackage[T1, T2A]{fontenc}
\usepackage[utf8]{inputenc}
\usepackage{graphicx}
\usepackage{booktabs}
\usepackage{amsmath}
\usepackage{xcolor}
\usepackage{multirow}
\usepackage{diagbox}
\usepackage{dirtytalk}
\usepackage{academicons}
\usepackage{colortbl}
\usepackage[most]{tcolorbox}
\usepackage{enumitem}
\usepackage{xurl}
\usepackage{cite}
\usepackage{csquotes}
\usepackage{listings}
\usepackage[title]{appendix}
\usepackage{tablefootnote}
\usepackage{algorithm}
\usepackage{algpseudocode}
\usepackage{epigraph}
\usepackage{booktabs}
\usepackage{subfig}
\usepackage{shapepar}
\usepackage{float}
\usepackage[depth=3]{bookmark}
\usepackage{minitoc}
\usepackage{pifont}
\usepackage{orcidlink}
\usepackage{bm}
\usepackage{tikz}
\usepackage{tikz-imagelabels} 

\definecolor{adversarial}{HTML}{971a1e}
\definecolor{color_a}{HTML}{893244}
\definecolor{color_b}{HTML}{4F404C}
\definecolor{color_c}{HTML}{007F7F}
\definecolor{color_d}{HTML}{6082B6}

\urldef{\UnicodeUtilitiesConfusables}\url{https://util.unicode.org/UnicodeJsps/confusables.jsp?a=poison&r=None}
\urldef{\MisspelledEnglishWordsMachine}\url{https://en.wikipedia.org/wiki/Wikipedia:Lists_of_common_misspellings/For_machines}
\urldef{\MisspelledEnglishWordsHuman}\url{https://en.wikipedia.org/wiki/Commonly_misspelled_English_words}
\urldef{\InvisibleCharacters}\url{https://invisible-characters.com/}
\urldef{\Wurzburg}\url{https://search.r-project.org/CRAN/refmans/stylo/html/dist.wurzburg.html}
\urldef{\Minmax}\url{https://search.r-project.org/CRAN/refmans/stylo/html/dist.minmax.html}
\urldef{\KagiTranslate}\url{https://translate.kagi.com/}

\imagelabelset{
    annotation arrow/.style = {
        preaction = {
            draw,
            -{Circle[fill=white, length=\tipsize+2*\borderthickness,
            width=\tipsize+2*\borderthickness]},
            line width = 2*\borderthickness + \arrowthickness,
            white,
            shorten >= \arrowdistance,
        },
        draw,
        -{Circle[fill=adversarial, length=\tipsize, width=\tipsize]},
        adversarial,
        line width = \arrowthickness,
        shorten >= \borderthickness + \arrowdistance,
    },
    annotation node/.style = {
        font=\annotationfont,
        inner sep = 0.5ex,
    },
    annotation font = \normalfont\tiny,
    tip size = 1mm
}

\title{
    Hijacking Text Heritage: Hiding the Human Signature through Homoglyphic Substitution
}

\author{
    Robert Dilworth \orcidlink{0009-0005-5497-9810}
}

\authorrunning{
    Robert Dilworth
}
\titlerunning{
    Doppelg\"anger Injection
}

\institute{
    Department of Computer Science and Engineering, Mississippi State University, Mississippi State, Mississippi, USA\\
    \email{rkd103@msstate.edu}
}

\hypersetup{
    pdftitle={Hijacking Text Heritage: Hiding the Human Signature through Homoglyphic Substitution},
    pdfsubject={cs.CR, cs.CL, cs.IR},
    pdfauthor={Robert Dilworth},
    pdfkeywords={Privacy, Adversarial Stylometry, Stylometry, Steganography, Zero-width Unicode Characters, Homoglyph Substitution, Translation, Obfuscation, Imitation, Injection, Authorship Verification, Authorship Attribution, TraceTarnish},
    colorlinks=true,
    linkcolor=color_a,
    citecolor=color_c,    
    urlcolor=color_d,
}

\begin{document}

\maketitle

\begin{abstract}

    In what way could a data breach involving government-issued IDs such as passports, driver's licenses, etc., rival a random voluntary disclosure on a nondescript social-media platform? At first glance, the former appears more significant, and that is a valid assessment. The disclosed data could contain an individual's date of birth and address; for all intents and purposes, a leak of that data would be disastrous. Given the threat, the latter scenario involving an innocuous online post seems comparatively harmless---or does it? From that post and others like it, a forensic linguist could stylometrically uncover equivalent pieces of information, estimating an age range for the author (adolescent or adult) and narrowing down their geographical location (specific country). While not an exact science---the determinations are statistical---stylometry can reveal comparable, though noticeably diluted, information about an individual. To prevent an ID from being breached, simply sharing it as little as possible suffices. Preventing the leakage of personal information from written text requires a more complex solution: adversarial stylometry. In this paper, we explore how performing homoglyph substitution---the replacement of characters with visually similar alternatives (e.g., ``h'' \texttt{[U+0068]} \( \rightarrow \) ``һ'' \texttt{[U+04BB]})---on text can degrade stylometric systems.

    
\end{abstract}

\section{Introduction}
\label{sec:Introduction}

    \epigraph{\textcolor{adversarial}{2 + ``2'' = 5}}{\textit{Nineteen Eighty-Four \\ George Orwell}}

    In this work, our goal is to measure the stylometric effect of incremental Injection of homoglyphs into a body of text. By running this experiment, we hope to quantify the efficacy of Injection---an adversarial stylometry technique explored in our prior work (\textit{Dilworth} \cite{UnicodeAdversarialStylometry2025,TraceTarnish2025,StegoStylo2026})---through homoglyphic substitution. The end goal, pending the outcome of our experiment, is the incorporation of a previously proposed technique---Injection by steganographic embedding of zero-width Unicode characters---with the current object of study---Injection by homoglyphic substitution. 
    
    Much like our previous research, the idea is to find the optimal level of Injection to induce authorship obfuscation, bastardize, or otherwise poison authorship attribution and verification systems. Given our undertaking, it is worth prefacing that, \textit{yes}, there exist beneficent use cases for text stylometry\footnote{All references to ``stylometry'' in this work pertain specifically to text stylometry, and any mention of ``steganography'' should likewise be understood as referring to text steganography.}. While not necessarily a destructive implement, stylometry \cite{Stropkay2025} is, at the end of the day, nothing but a tool. The assumption here is that the tool can be, or has already become, a weapon. As such, whenever a tool becomes a weapon---as we suggest is the case with stylometry---preventive measures become logically unavoidable.

    As a preliminary primer on homoglyphic substitution, which we will explore further in later sections, consider a simple mathematical expression, for example: two plus two equals five. Based on elementary schooling, this expression is unequivocally false, but with enough technical equivocation, it can be made to appear valid. The mechanism---or the confounding element that can render a falsity a truth---that could be introduced into the system is the concept of a variable, a placeholder of sorts. If we designate a placeholder that, for all intents and purposes, resembles the number two but holds the numerical value of three, we can construct an argument proving the statement's validity. In this scenario, our proof would begin with the argument that the symbol representing the number two can either denote the presence of two objects or serve as a placeholder for another numerical value. This ambiguity suffices for our example, but it poses the risk of logical errors when the proof's consequences are extrapolated further. This nebulousness and the uncertainty it spawns are the crux of homoglyph attacks, albeit in a simplified form. In practice, such attacks appear in phishing schemes, where the URL of a reputable service is spoofed by replacing characters with similar-looking but differently interpreted ones, redirecting traffic to actor-controlled websites and domains. A common outcome of these attacks is credential harvesting, which can potentially lead to privilege escalation.

    Having said thıs\footnote{To illustrate a straightforward example of homoglyphic substitution, we have replaced the lowercase ``i'' \texttt{[U+0069]} in ``this'' with the Latin small letter dotless ``\i'' \texttt{[U+0131]}.}, a clear distinction must be made between the contemporary definition of a homoglyph attack and our implementation, for the distinction between a tool and a weapon can blur at a moment's notice. Our motivation for engineering what some may describe as a homoglyph attack is privacy. The scope of the attack is to target a specific system---stylometric systems---and nothing more. The same cannot be said of the more pernicious variant that ultimately impinges upon privacy. For the sake of desperately clinging to the vestiges of privacy that have slowly eroded with time, we continue our probe into the tactics, tools, and procedures best suited to hampering stylometric analysis---contaminate (\textit{poison the well}), catalyze (\textit{feed the fire}), and confound (\textit{douse the blaze with bane}).

    \subsection{Paper Structure}
    \label{subsec:Paper_Structure}

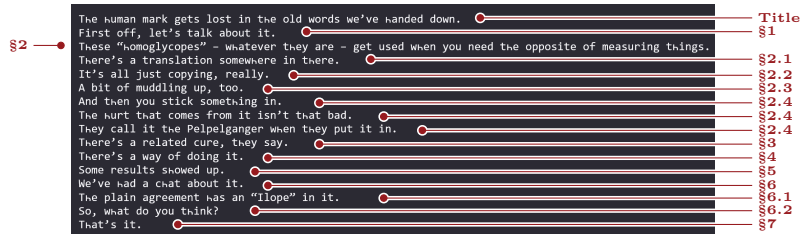
\begin{figure}[H]
            \centering
            \begin{annotationimage}{width=0.7\textwidth}{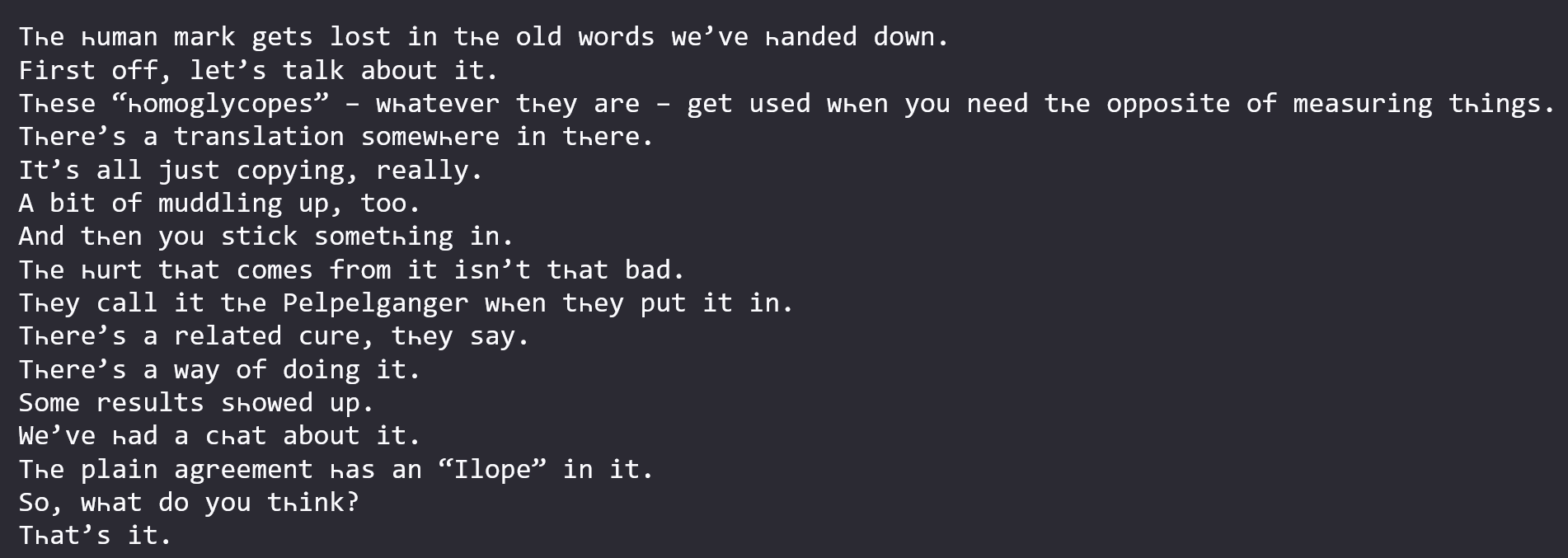}
                \draw[annotation right = {\textbf{Title} at 0.94}] to (0.61, 0.94);
                \draw[annotation right = {\textbf{\S \ref{sec:Introduction}} at 0.88}] to (0.34, 0.88);
                \draw[annotation left = {\textbf{\S \ref{sec:Why_Use_Homoglyphs}} at 0.82}] to (0.01, 0.82);
                \draw[annotation right = {\textbf{\S \ref{subsec:Translation}} at 0.76}] to (0.44, 0.76);
                \draw[annotation right = {\textbf{\S \ref{subsec:Imitation}} at 0.69}] to (0.32, 0.69);
                \draw[annotation right = {\textbf{\S \ref{subsec:Obfuscation}} at 0.63}] to (0.28, 0.63);
                \draw[annotation right = {\textbf{\S \ref{subsec:Injection}} at 0.57}] to (0.36, 0.57);
                \draw[annotation right = {\textbf{\S \ref{subsubsec:Liminal_Injection}} at 0.51}] to (0.46, 0.51);
                \draw[annotation right = {\textbf{\S \ref{subsubsec:Doppelganger_Injection}} at 0.45}] to (0.52, 0.45);
                \draw[annotation right = {\textbf{\S \ref{sec:Related_Work}} at 0.39}] to (0.36, 0.39);
                \draw[annotation right = {\textbf{\S \ref{sec:Methodology}} at 0.33}] to (0.28, 0.33);
                \draw[annotation right = {\textbf{\S \ref{sec:Results}} at 0.27}] to (0.26, 0.27);
                \draw[annotation right = {\textbf{\S \ref{sec:Discussion}} at 0.21}] to (0.28, 0.21);
                \draw[annotation right = {\textbf{\S \ref{subsec:The_Illusion}} at 0.155}] to (0.46, 0.155);
                \draw[annotation right = {\textbf{\S \ref{subsec:Beyond_the_Fine}} at 0.1}] to (0.26, 0.1);
                \draw[annotation right = {\textbf{\S \ref{sec:Conclusion}} at 0.04}] to (0.14, 0.04);
            \end{annotationimage}
            \caption{The paper's structure (with annotations) post \textsc{TraceTarnish} processing. \S\S \ref{subsubsec:LLM_Translate_Imitation}, \ref{subsubsec:Surrealist_Injection}, \ref{subsec:Privacy_Paradox}, \ref{subsec:Reading_Between}, \ref{subsec:Stylometric_Fingerprints}, and \ref{subsec:Data_Poisoning} were not processed because their exclusion produced a more cohesive output.}
            \label{fig:Paper_Structure}
        \end{figure}

        The remainder of the paper is organized as follows, albeit in a rather unconventional fashion. To demonstrate our \textsc{TraceTarnish} attack, we altered the section headings of this paper (excluding the section that describes the paper's structure) to conceal authorship and mask the fact that we authored the text. The resulting content has undergone a fourfold transformation, the details of which we describe later in the paper. The takeaway is that, unless one actively searches for it, the \textit{traces} of the attack remain perceptibly obscured, hiding the \textit{tarnish} beneath. See (\textbf{Figure \ref{fig:Paper_Structure}}).
        
\section{Why Use Homoglyphs for Adversarial Stylometry}
\label{sec:Why_Use_Homoglyphs}

    \epigraph{\textcolor{adversarial}{Even to understand the word `doublethink' involved the use of doublethink.}}{\textit{Nineteen Eighty-Four \\ George Orwell}}

    In hindsight, our attempt to circumvent stylometric analysis inadvertently led to a non-exhaustive search of viable methods to achieve that goal. Reviewing the foundational literature on adversarial stylometry reveals three established approaches: Translation, Imitation, and Obfuscation.

    \subsection{Translation}
    \label{subsec:Translation}

        Putting aside recent advancements in machine learning, such as Google's Neural Machine Translation, the adversarial effect of performing multiple translations heavily relies on the natural loss of meaning inherent to language conversion. Consider the many examples of loanwords that have been borrowed between languages when a suitable, comparable word does not exist in the lexicons \cite{Keswani2016}.

        In this context, \textbf{back-translation}---or \textbf{roundtrip translation}---illustrates how translating a text back and forth between languages can lead to further distortions. Each translation phase may introduce new nuances or misinterpretations, compounding the original loss of meaning. This demonstrates the fragility of linguistic fidelity in the face of mechanical transitions across languages.

    \subsection{Imitation}
    \label{subsec:Imitation}

        Imitation, historically speaking, is as old as humanity itself. From the high-profile forgeries of renowned artists to the replication of culturally significant artifacts, copying a creation is a natural part of the human condition. Its deployment in this domain is no different, although the incentives may vary: transferring the style of author \( A \) to work \( B \) while retaining \( B \)'s meaning and masking \( B \)'s stylistic cues.

        \subsubsection{Using an LLM-Powered Translator to Actualize ``Imitative Translation.''}
        \label{subsubsec:LLM_Translate_Imitation}

            Despite its name, Kagi Translate\footnote{\KagiTranslate}, like its competitors, offers an interface for converting text from one language to another. What sets it apart and merits our interest is one of its AI-enabled features: the translator allows the user to specify the source or target language explicitly, regardless of whether the target or source language is an actual language. This capability, if taken at face value, can be leveraged in an adversarial stylometry workflow as a method of Imitation.

            Based on the description of the tool and its intended purpose, that may seem counterintuitive. Accordingly, we propose two ways in which this tool could foster privacy in text-based communications.

            As we established, Kagi Translate permits the user to specify the target language. If you supply a sensible input, such as \say{Spanish,} the translator will function as designed: it will convert the source text---whatever language it may be written in---into Spanish. However, what if we were to enter a target language of \say{George Orwell}? What would the expected output be? Based on our observations, the resulting text would imitate the writing (or speaking) style of George Orwell, more or less preserving the underlying meaning while rendering it in a manner reminiscent of the author.

            Other amusing possibilities include having your text appear as if authored by various personas, assuming accents, taking on the guise of poets or musicians, or emulating speaking characteristics native to certain social-media spheres. Assuming the underlying Large Language Model (LLM) has been trained on suitable material, the possibilities for what the target \say{language} can be seem limitless\footnote{It remains to be seen whether the data used to train the model were ethically sourced.}.

            It is in this seemingly infinite expanse that the wheels in our heads began to turn. Suppose you wish to send a message anonymously and have it translated into a target language with personality or stylistic qualifiers---say, into German, but with the author's personality \say{\textit{a},} demographics \say{\textit{b}, \textit{c}, \textit{d},} age \say{\textit{e},} etc. What if you chained a series of such personalized translations, randomizing the target language and the personality qualifiers that influence the text's stylistic composition? After a series of such translations, would the resulting text, when translated back into the desired source language, retain the original's stylistic cues?

            From a purely imitative perspective, there is similar potential: take text written in your desired source language and have Kagi Translate render it into a target \say{language} that is a well-known individual, such as an author. The same question arises: would the author's stylistic fingerprint be overwritten by Kagi Translate as it translates to the specified target language?

\begin{figure}[H]
                \centering
                \includegraphics[width=1\linewidth]{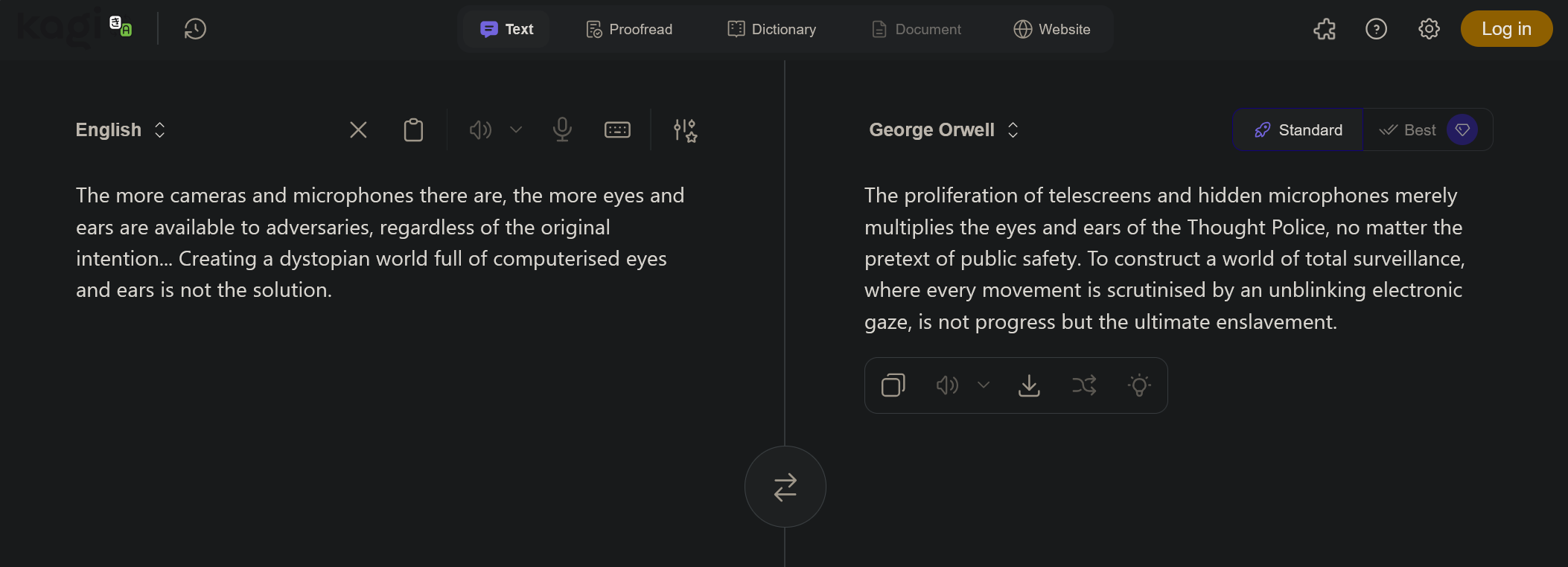}
                \caption{Kagi Translate acts as a conduit for adversarial stylometry, rending authorship asunder when it is used for ``Imitative Translation'' (translating a text while trying to mimic another author's style). The target ``language'' for the translation is ``George Orwell.''}
                \label{fig:Kagi_Translate_Imitative_Translation_Example}
            \end{figure}

            While this was perhaps not the developer's intent for the service, the translator may be a convenient and simple tool for adversarial stylometry, especially if the output of the Imitative Translation undergoes the Injection regimen we will discuss later. For lucidity, \textbf{Imitative Translation} is the process of using an LLM-powered translator to render text in the stylistic voice of a specified persona or author rather than a conventional language. See (\textbf{Figure \ref{fig:Kagi_Translate_Imitative_Translation_Example}}) for an example of Imitative Translation.

    \subsection{Obfuscation}
    \label{subsec:Obfuscation}

        Obfuscation, in a linguistic sense---and according to the literature---is usually achieved through \textbf{paraphrasing}. Restating an original thought in a more concise, less bloated form fundamentally trims excess language, eliminating florid expressions, leaving only a plain, terse rendition. This transformation marks the adversarial effect of obfuscation.

        Similarly, \textbf{synonym substitution}, the replacement of words and phrases with synonymous words and equivalents, serves a parallel function. By exchanging complex or obscure terms for simpler, more familiar ones, it creates clarity and accessibility, suppressing linguistic signals of the original text. This technique effectively maintains the original meaning while reinforcing the essence of obfuscation.

    \subsection{Injection}
    \label{subsec:Injection}

        As we developed an attack---\textsc{TraceTarnish}---leveraging the strategies of Translation, Imitation, and Obfuscation, we formulated a novel approach, which we have dubbed \say{Injection.} From a technical standpoint, the process itself is not innovative, but the method of achieving the desired adversarial effect arguably is.

        As the name suggests, the approach can be encapsulated straightforwardly: either interpolate or supplant text.

        \subsubsection{Liminal Injection.}
        \label{subsubsec:Liminal_Injection}

            Our approach defines the first variant: Liminal Injection, the insertion of zero-width Unicode characters via steganography\footnote{An illustrative application of steganography employing invisible Unicode characters is presented in \textit{Sundar} \cite{Sundar2020}. The application, \textit{StegCloak}, converts a plaintext message and the password used to access that message into ciphertext using salted encryption, encoding the ciphertext with zero-width Unicode characters in a way that can be easily embedded in cover text (the text meant to innocuously house the stego text). A noteworthy finding from the write-up is that, despite the expansive list of invisible characters at our disposal, deploying them in the wild and subjecting them to various platform blocks and blacklisting leaves only a remnant to exploit: zero-width non-joiner \texttt{[U+200C]} and zero-width joiner \texttt{[U+200D]}.} \say{in-between} the characters of words in a sentence. For example, consider a sentence \( S \) containing words \( W_{0}, W_{1}, \dots, W_{13} \). A Liminal Injection attack---also referred to as an \textit{encoding attack} (\textit{Mosquera} \cite{Mosquera2022}) or \textit{invisible-character attack} (\textit{Teja et al.} \cite{Teja2026})---evaluates the individual letters \( (l_{0}, l_{1}, \dots, l_{11}) \) within each word and inserts a zero-width Unicode character---space, non-joiner, joiner, or no-break space\footnote{A list of invisible, non-printable Unicode characters is available here: \InvisibleCharacters.}---after the first letter \( l_{0} \).

            In our testing, the placement of the zero-width Unicode character appears largely inconsequential; we chose that location merely for convenience. The primary motivation for our placement is to avoid data sanitization. Leading and trailing string sanitization---i.e., text cleansing that occurs immediately before and after a word---is common, but this attack remains viable when less stringent scrubbing standards are used. So, if positions \( l_{0} \) or \( l_{n} \) (the first and last characters) are populated with zero-width Unicode characters, even the lowest common denominator of security practices---provided they are applied at all---should be able to detect the tampering\footnote{More robust text-perturbation defenses are proposed by \textit{Bhalerao et al.} \cite{Bhalerao2022}.}.

            While effective when the recipient or processor is unaware of its existence, what happens if a service applies more diligent data sanitization? In that case, the second variant---Doppelg\"anger Injection---can replicate the clandestine adversarial effect of the first while relying on easily mistaken characters or homoglyphs.

        \subsubsection{Doppelg\"anger Injection.}
        \label{subsubsec:Doppelganger_Injection}

            It is said that all living creatures have an identical \textit{double}, or second self, that concurrently inhabits and roams the earth. Doppelg\"angers---a German loanword used to describe such counterparts---often resemble a creature superficially but differ from it in a deep-seated way. The same could be said of homoglyphs, but replace the doppelg\"anger \say{creature} with symbolic markers (\say{characters}) and the antipodal \say{evil twin} connotation with \say{contrasting typographical meaning.}

            Put another way, and setting aside the discussion of doppelg\"angers, a homoglyph is a character that superficially resembles another character but differs in linguistic meaning. An example of Doppelg\"anger Injection would be taking a word \( W \) and replacing a variable number of its letters---\( l_{0}, l_{1}, \dots, l_{7} \)---with homoglyphs \( h_{0}, h_{1}, \dots, h_{n} \). Such tampering should conceivably induce the same adversarial effect, since stylometry, reductively speaking, relies on character matching. Introducing mismatches, whether with zero-width Unicode characters or homoglyphs\footnote{\textit{Font-based perturbation} (\textit{Zhang et al.} \cite{Zhang2025})---a closely related attack adjacent to homoglyph attacks---uses special fonts that are not necessarily homoglyphs but serve more broadly as decorative elements within text, such as font-like emoji. These expressive tools---think regional indicator symbols, mathematical alphabets, circled letters, and squared letters---exploit the same human-machine perception gap that homoglyphs prey upon. The stylistic fonts, not inherently tied to another language like the Cyrillic alphabet, induce a similar adversarial effect on text. Substituting a character with a visually similar stylistic font perturbs text processing, such as tokenization. Based on this logic, such an attack should also negatively impact a stylometric system, although the poisoned artifacts would be extremely distinct by the very nature of the attack.}, should help cloak authorship.

        \subsubsection{Surrealist Injection.}
        \label{subsubsec:Surrealist_Injection}

            A picture is worth a thousand words, but in the case of the surrealist painter Salvador Dal\'{\i}, the complexity he conveyed was exponential. A single piece of his work presents an expansive domain rife with topsy-turvy, illogical imagery, challenging the rational and natural: fantastically incongruous, irrationally juxtaposed, symbolic, subversive, outlandish, peculiar, and uncanny. Such descriptors only scratch the surface of not only the artist himself but also the movement with which he is so closely associated.

            The eccentric essence of Dal\'{\i} is what serves as the namesake for this Injection variant. The underlying question, then, is: what is the most jarring thing that could be encountered in a written piece of text, barring the inclusion of provocative linguistic constructs, sensational media, or unorthodox formatting? 

            On the whole, to communicate an idea, the underlying words or tokens comprising the message must be recognizable. So, as a matter of course, the way words are spelled holds great consequence, and the origin of such misspellings may spring from typographical or phonological errors. What is being intentionally instilled within text is the surrealist equivalent of a \say{strange element} betraying expectation and conformity. In painting, this could materialize as a melting clock; in writing, creative or accidental misspelling serves that role (as comparatively uncaptivating as that may be).

            On a technical level, Surrealist Injection scans the words within a body of text, replacing certain words with typographical or phonological misspellings. Take a word \( W \) and corrupt it such that either (A) it appears as if the wrong sequence of characters were typed to produce it, or (B) the wrong sequence of sounds underlying the word (phonemes) were mistaken, producing a similar stream of invalid characters (\textit{Rumpf} \cite{Rumpf}). Here, the \say{undirected play of thought} and the \say{negation of the conventional} herald a break from tradition, with tradition being the rules governing spelling.

\begin{figure}[H]
            \centering
            \subfloat{\includegraphics[width=0.45\linewidth]{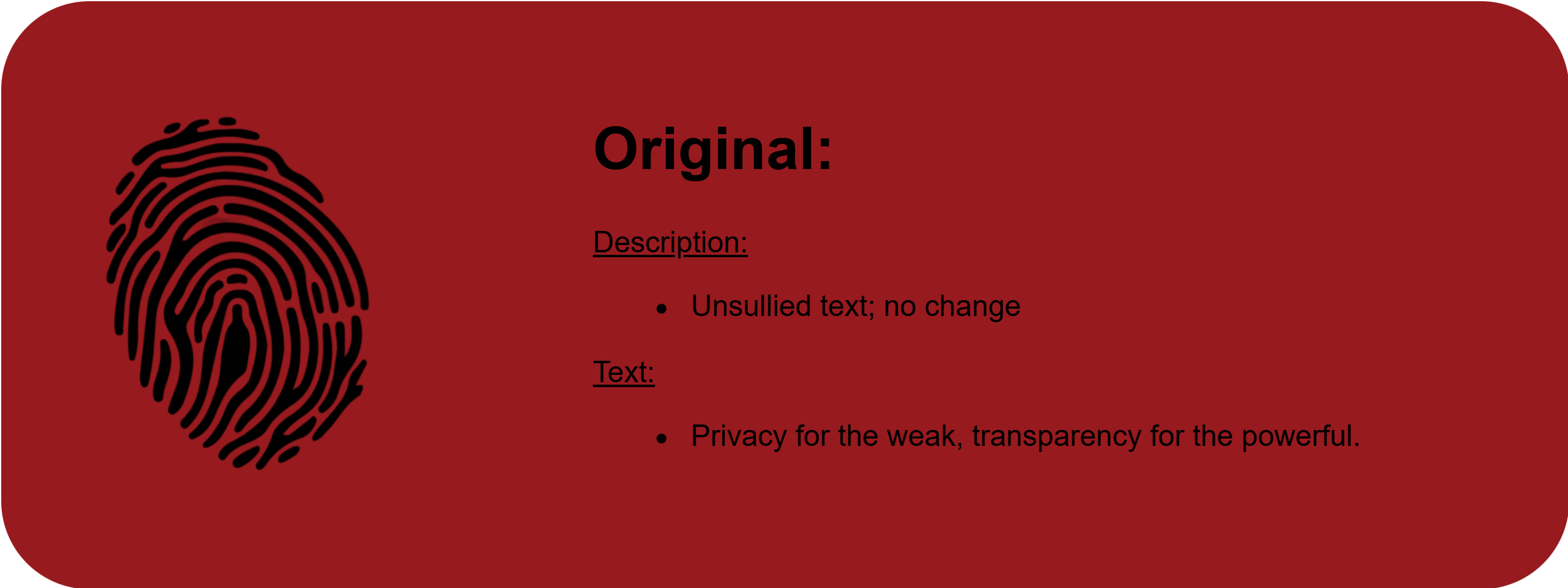}} \quad
            \subfloat{\includegraphics[width=0.45\linewidth]{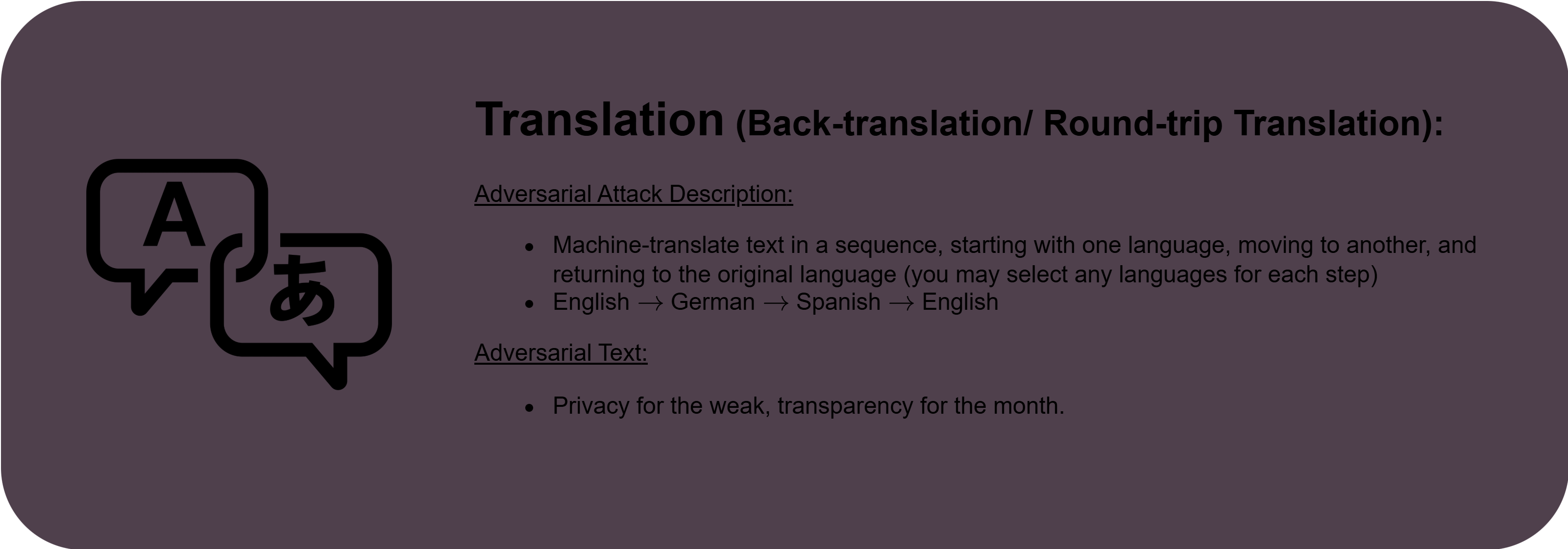}} \quad
            \subfloat{\includegraphics[width=0.45\linewidth]{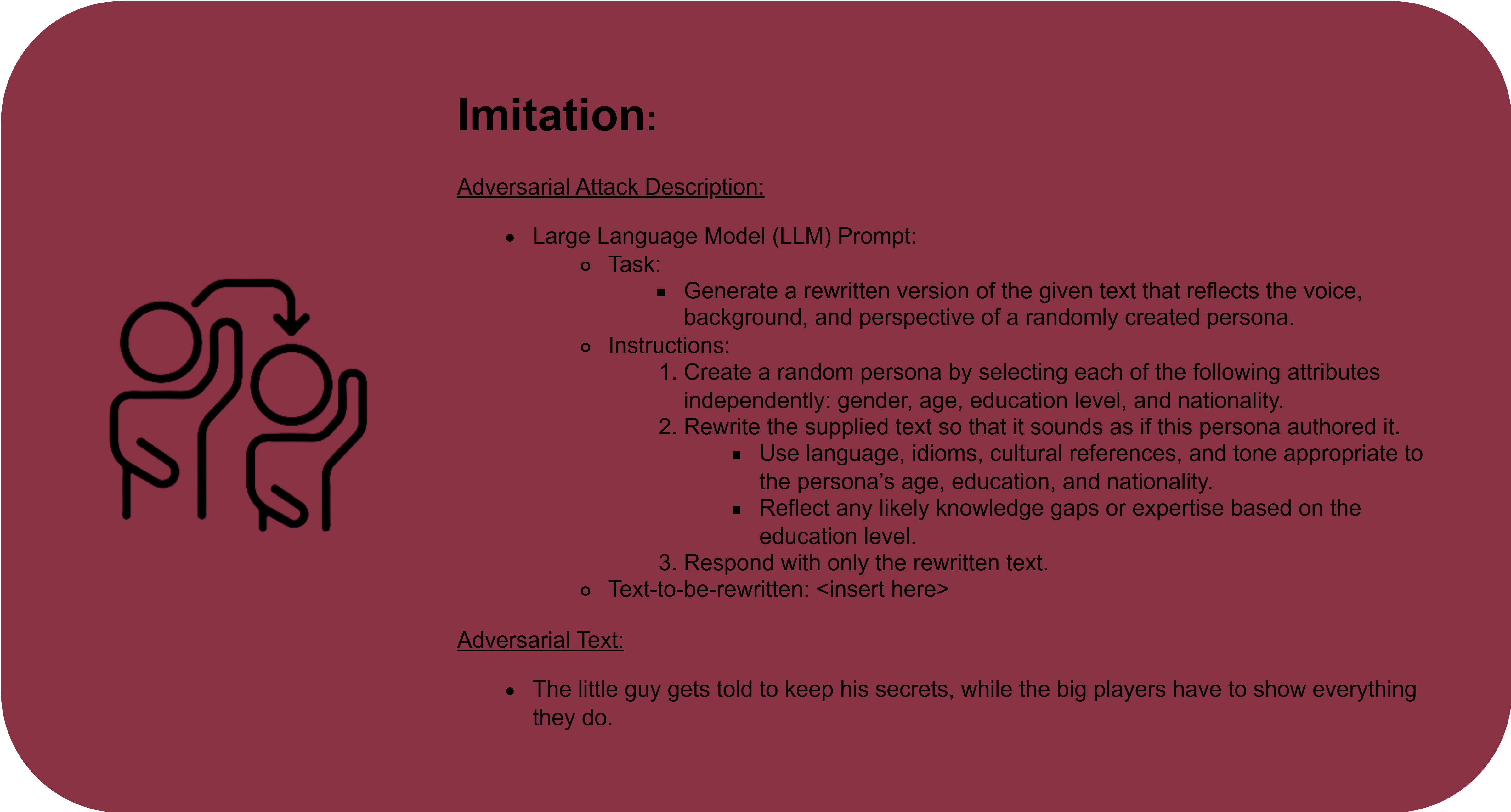}} \quad
            \subfloat{\includegraphics[width=0.45\linewidth]{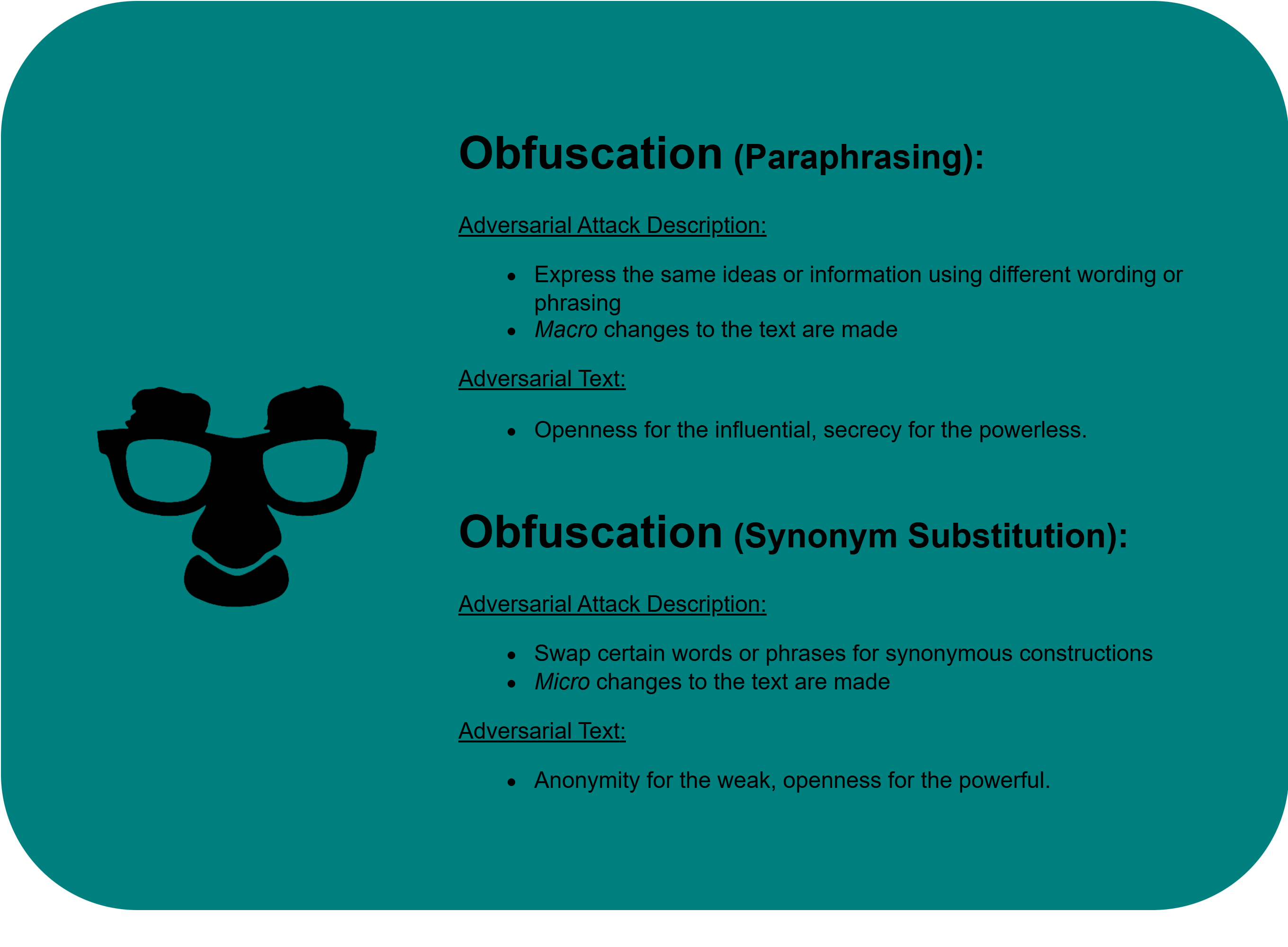}} \quad
            \subfloat{\includegraphics[width=0.45\linewidth]{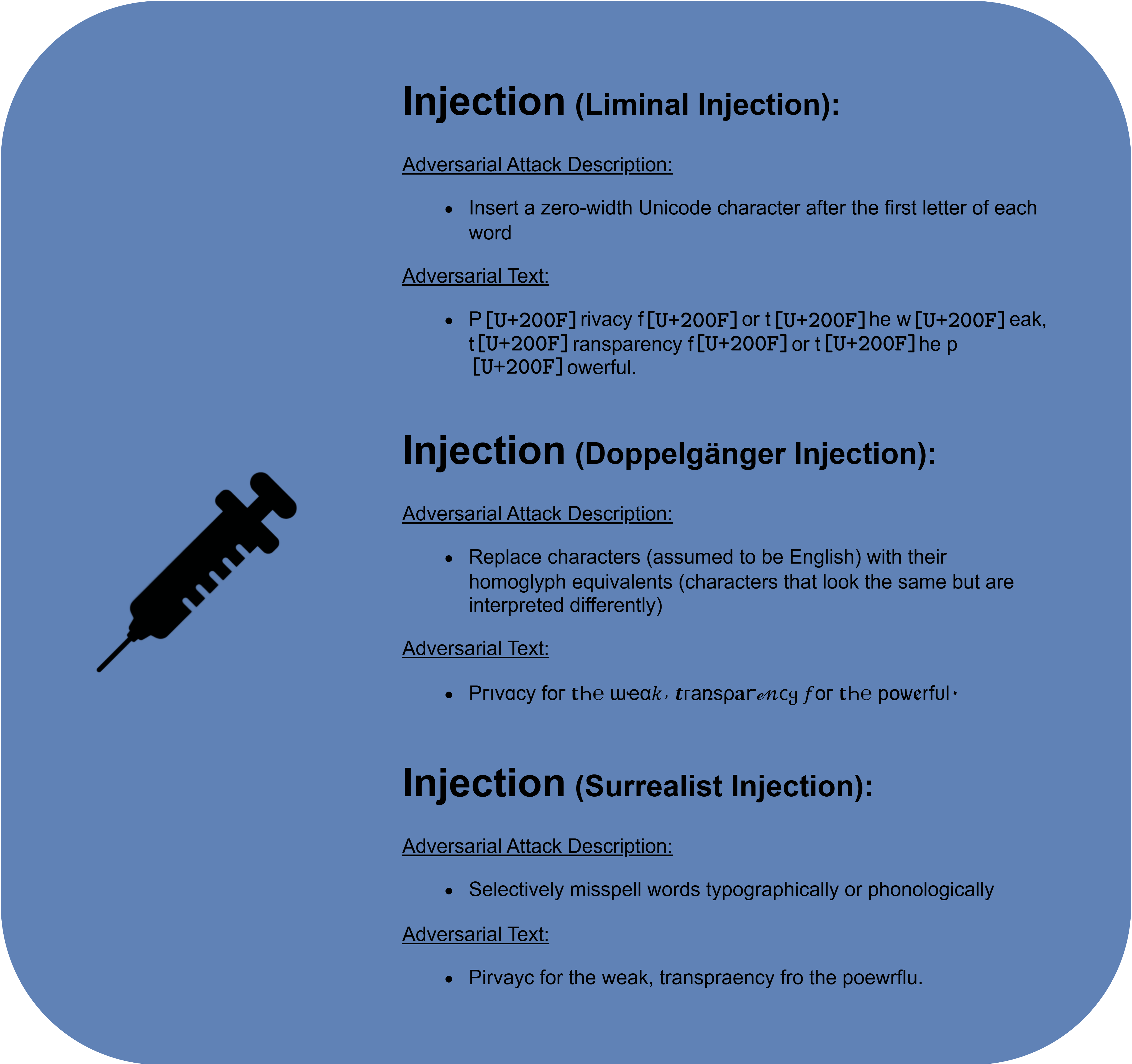}} \quad
            \subfloat{\includegraphics[width=0.45\linewidth]{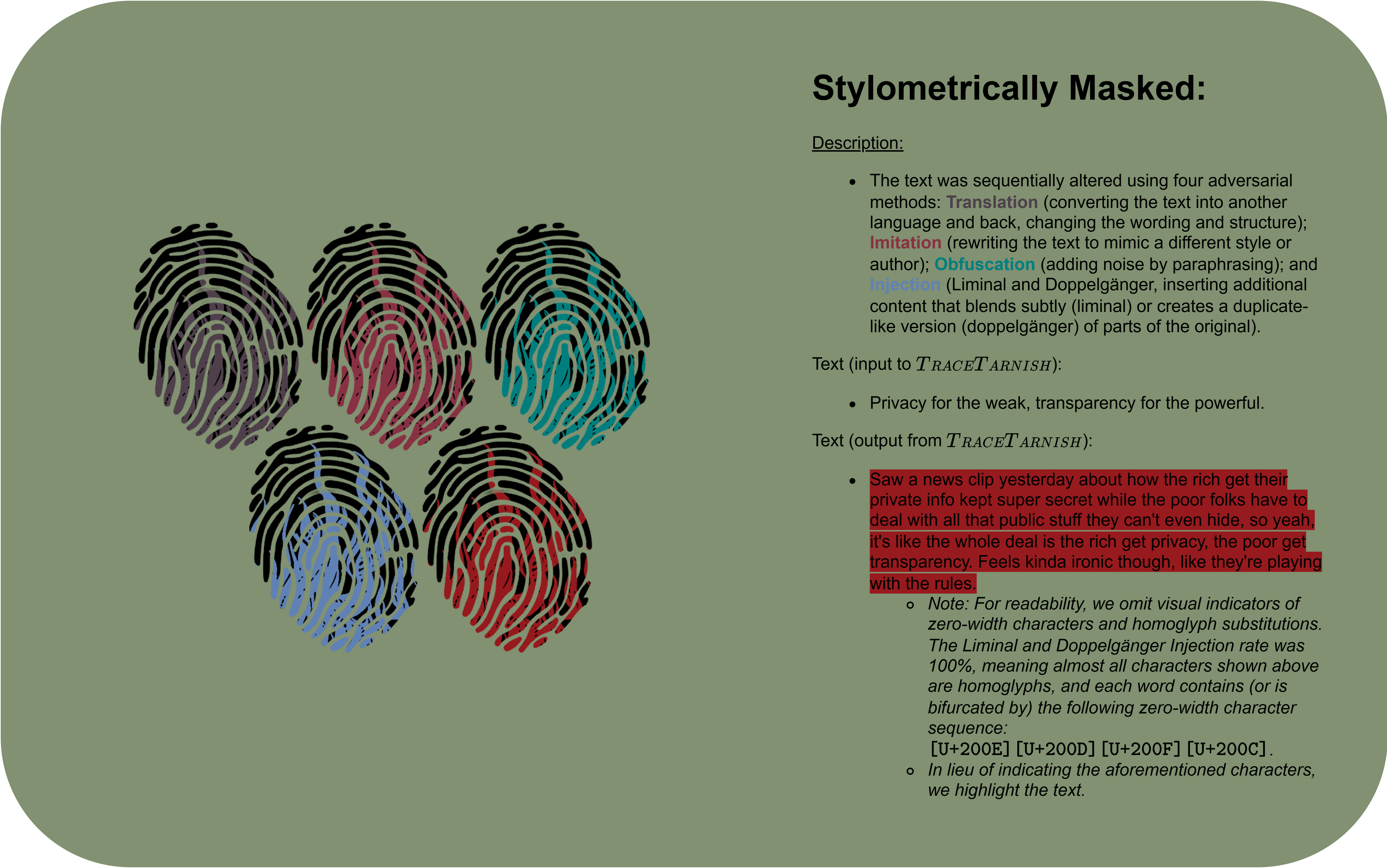}}
            \caption{A Taxonomic Overview of the Adversarial Attacks: \textcolor{color_b}{Translation}, \textcolor{color_a}{Imitation}, \textcolor{color_c}{Obfuscation}, and \textcolor{color_d}{Injection}}
            \label{fig:Adversarial_Stylometry_Taxonomy}
        \end{figure}

\begin{figure}[H]
            \centering
            \includegraphics[width=1\linewidth]{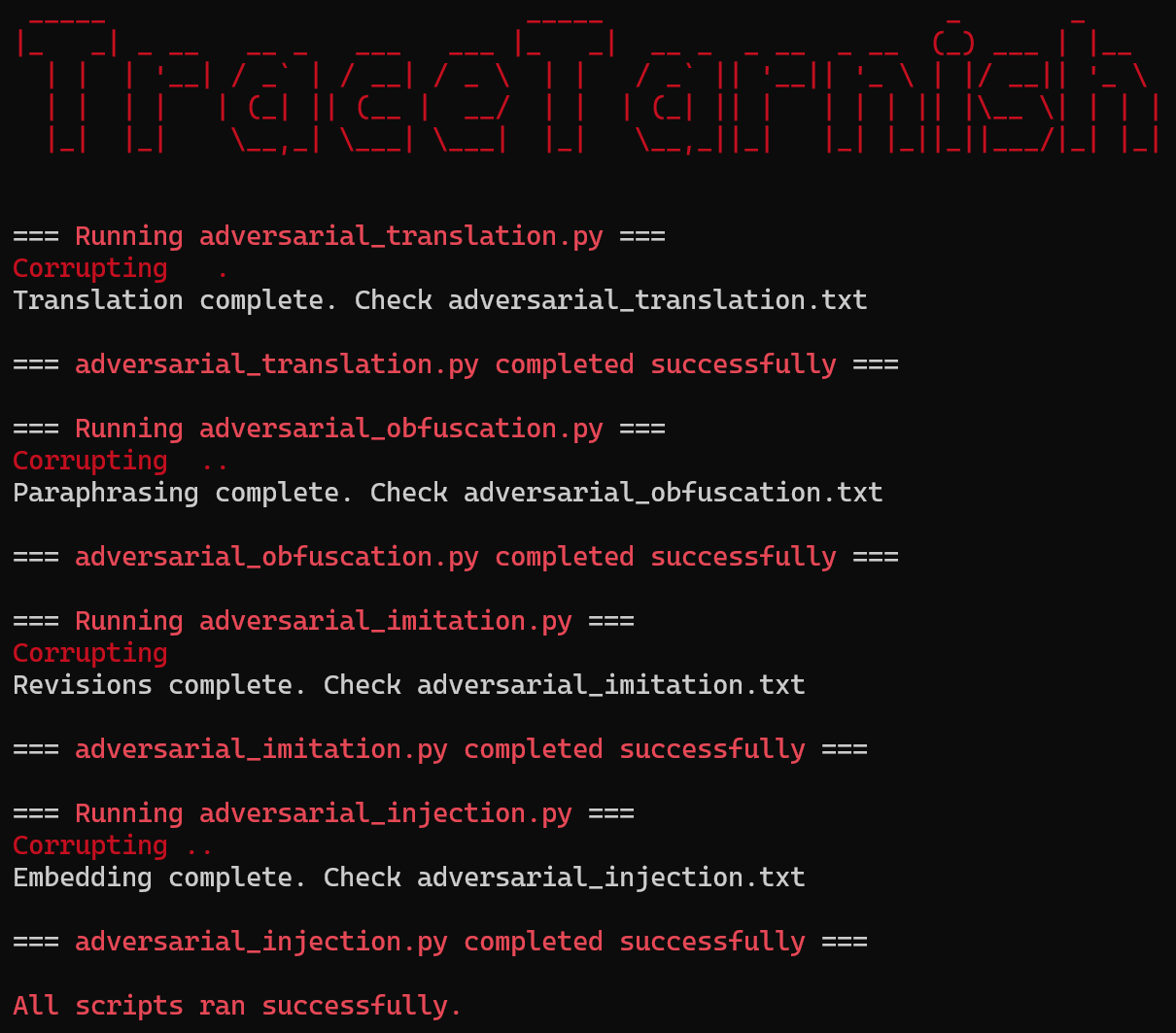}
            \caption{\textsc{TraceTarnish}: Our stylometric attack script---a gestalt modular framework where each component contributes to a whole that is greater than the sum of its parts; incorporating homoglyph functionality resulted in the following processing pipeline for razing authorship: \textcolor{color_b}{Translation} \( \rightarrow \) \textcolor{color_c}{Obfuscation} \( \rightarrow \) \textcolor{color_a}{Imitation} \( \rightarrow \) \textcolor{color_d}{Injection}}
            \label{fig:TraceTarnish_v3_Terminal_Output}
        \end{figure}

        To the best of our knowledge, and considering existing strategies---Translation, Imitation, Obfuscation, and Injection (Liminal, Doppelg\"anger, and Surrealist)---no study has evaluated the effectiveness of Doppelg\"anger Injection (or homoglyph-based attacks) in an adversarial stylometry setting. We aim to fill this gap with this paper. See (\textbf{Figure \ref{fig:Adversarial_Stylometry_Taxonomy}}) for a taxonomic overview of the adversarial attacks discussed above. See (\textbf{Figure \ref{fig:TraceTarnish_v3_Terminal_Output}}) for the current implementation of our attack script, \textsc{TraceTarnish}, which makes use of the strategies shown in (\textbf{Figure \ref{fig:Adversarial_Stylometry_Taxonomy}}).

\section{Related Work}
\label{sec:Related_Work}

    \epigraph{\textcolor{adversarial}{Nothing holds [the Brotherhood] together except an idea which is indestructible. You will never have anything to sustain you, except the idea. You will get no comradeship and no encouragement. When finally you are caught, you will get no help. You will have to get used to living without results and without hope. You will work\dots, you will be caught, you will confess, and then you will die\dots We are the dead.}}{\textit{Nineteen Eighty-Four \\ George Orwell}}

    In this section, we peruse publications on Unicode \textit{confusables}\footnote{A Unicode confusable is a character that looks visually identical or very similar to another character but has a different code point. A list of visually confusable characters can be found here: \UnicodeUtilitiesConfusables.}, such as homoglyphs, and examine their applications in fields related to our study, from evading machine-generated text detectors to various other aims.

    \textbf{\textit{Creo et al.} \cite{Creo2025}} evaluate the effectiveness of homoglyph-based attacks on AI-generated text detectors, applying attacks with varying replacement percentages---5\%, 10\%, 15\%, and 20\%---to randomly chosen characters or to all replaceable characters, which they refer to as a \say{greedy attack.}

    The task of distinguishing human-written text from AI-generated text is similar to, but distinct from, the task of attributing authorship among different authors. If the corpus of human-written text is assumed to have been produced by a single author (an admittedly incorrect assumption) and the AI-generated text is attributed to another sole author, then the findings can be applied to our study.

    Broadly speaking, because homoglyphs have different encodings, tokenizers treat them differently. In this context, attack tokens---text altered by a homoglyph attack---are split into smaller tokens by an AI detection system, increasing the total number of tokens, \( N \). The researchers found that modifying 10\% of the characters changes the tokenization 70\% of the time. The authors conclude that homoglyph-based attacks effectively evade AI-generated text detection by inducing misclassifications. The presence of homoglyphs makes the output embeddings less discriminative, as the model cannot fully capture the text's semantics, thereby altering tokenization. The accompanying code can be used to perform homoglyph-based attacks, and we hope to measure its effect on authorship obfuscation and the thwarting of stylometric analysis.

    \textbf{\textit{Alvi et al.} \cite{Alvi2017}} devise methods to combat technically disguised plagiarism---plagiarism transformed through technical means, usually obfuscation, to evade detection. One way to potentially pass off plagiarized content as authentic is to change the computational representation of text, substituting characters with their homoglyphic equivalents. The reported failures of notable plagiarism-detection systems, which the paper aims to ameliorate, showcase the effectiveness of homoglyph-based attacks in this space. For instance, Turnitin appears resistant to homoglyph substitution, retaining its ability to locate sources for plagiarized content.  

    The researchers proposed two approaches for detecting plagiarism in homoglyph-obfuscated texts: the first uses the Unicode list of confusables to replace homoglyphs with a visually similar representation, while the second employs a similarity score computed using normalized Hamming distance.

    \textbf{\textit{Macko et al.} \cite{Macko2024}} evaluate the efficacy of machine-generated text detectors in the face of adversarial perturbations, implementing authorship-obfuscation techniques---back-translation, paraphrasing, and text edits---to masquerade detected text as undetected text. The class of text-edit authorship-obfuscation methods discussed in the paper is of most interest to our study, where the edits can be further subdivided into lexical, syntactic, morphological, or orthographic categories. From their experiments across multiple languages, the researchers found that homoglyph-based attacks have a 50\% or higher chance of successfully avoiding detection; however, such attacks are easily detected and remediated by preprocessing. The cited success of orthographic edits, or homoglyph attacks, lies in their ability to perturb word embeddings in the evaluated detection models.

    \textbf{\textit{Dugan et al.} \cite{Dugan2024}} proposed an extensive and robust benchmark dataset, RAID, to better appraise the success and scope of machine-generated text detection models, notably accounting for adversarial attacks to more accurately estimate and validate their capabilities. The research aims to facilitate the development of more accurate detectors to weed out spurious machine-generated text and mitigate their potentially deleterious effects. Of the eleven attacks introduced in the paper, we focus exclusively on the homoglyph-based attacks, which the researchers posited effectively diminished detection for all but one detector, GPTZero. On average, all other detectors degraded by 40.6\%, while GPTZero deteriorated less, by 0.3\%. 

    The paper also broaches the idea that attempting to attack an arbitrary detector with imperfect information about said detector would prove difficult, shedding light on the realities of our study. 

    In most circumstances, who has access to what data (in this case, an author's writings, online posts, etc.) and which stylometric system will be used to evaluate that data will almost always remain unknown, unless an author goes out of their way to discover an adversary's preferred system or methodology. In the absence of such crucial information---such as which detector or stylometric system an adversary will employ---and given the relative dearth of data, the paper utilized a homoglyph attack rate of 100\%, limiting substitutions to only those characters that are indistinguishable to the human eye, which facilitated the unrestrained homoglyph substitution of all possible characters.

    A major contribution of the work is its enumeration and explanation of the eleven adversarial attacks, which are: alternate spelling, article deletion, paragraph addition, case swapping, zero-width space insertion, whitespace augmentation, homoglyph substitution, digit shuffling, misspelling, paraphrasing, and synonym swapping. The strategies outlined therein are easily transferable to other domains, such as stylometric evasion and authorship obfuscation, which directly impact our study.

\begin{figure}[H]
        \centering
        \includegraphics[width=1\linewidth]{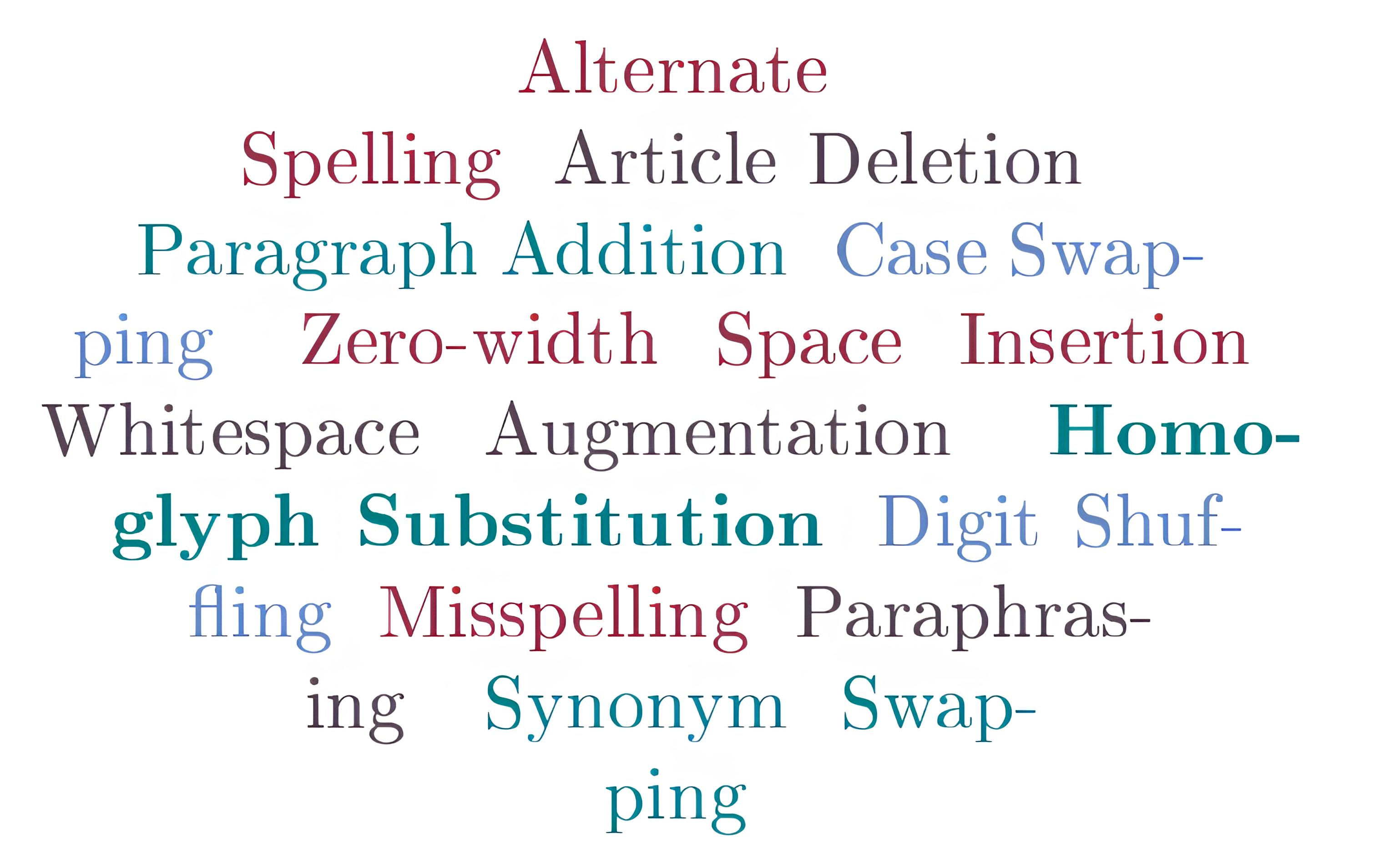}
        \caption{An enumeration of the adversarial attacks examined by \textit{Dugan et al.} \cite{Dugan2024}.}
        \label{fig:Adversarial_Attack_Enumeration}
    \end{figure}

    \textbf{\textit{Huang et al.} \cite{Huang2025}} provide a guidepost for future work in the adversarial stylometry space, presenting a template for adversarial attack avenues that can be further pursued and reiterating the findings of \textit{Dugan et al.} \cite{Dugan2024}; see (\textbf{Figure \ref{fig:Adversarial_Attack_Enumeration}}).

    \textbf{\textit{Gagiano et al.} \cite{Gagiano2021}} investigate the susceptibility of the GROVER model---a neural fake-news generator and discriminator---to adversarial attacks, particularly character-level and token-level perturbations. At the component-level, GROVER's generative functionality serves to effectuate and proliferate propaganda; its discriminative functionality works to curtail the dissemination of disinformation. 

    The researchers claim that GROVER, in its defensive stance, is highly vulnerable to adversarial attacks, failing to adequately differentiate between machine-generated and human-produced articles when the former undergo minor alterations (or adversarial attacks). The scrutinized attacks, as mentioned in the paper, were case swapping (uppercasing or lowercasing a letter), homoglyph substitution, whitespace augmentation (removing spaces), and misspelling. 

    Based on their findings, character-level attacks (such as homoglyph substitution) prove effective. With a misclassification rate of 4.39\%---second only to misspelling's 9.78\%---homoglyph substitution affected 97\% of the 100 sampled articles that were definitively classified as machine-generated by GROVER, flipping the classification from \say{Machine} to \say{Human.} To clarify, all 100 articles initially received a strong \say{Machine} classification, but after various iterations of the adversarial attacks, they shifted toward the \say{Human} classification. 

    An unexpected finding of this paper is the remarkable effectiveness with which a misspelling attack can induce misclassifications. Further exploration of the impact of misspelling attacks\footnote{Comprehensive lists of commonly misspelled English words that can be leveraged in a misspelling attack are available here (the machine-readable list: \MisspelledEnglishWordsMachine) and here (the human-readable list: \MisspelledEnglishWordsHuman).} on adversarial stylometry may be warranted\footnote{The character n-gram representation tolerates noisy text---grammatical errors or odd punctuation cause little impact. N-gram representations remain robust against adversarial misspellings intended to hide authorship; lexical models, as noted by \textit{Stamatatos} \cite{Stamatatos2009}, are vulnerable to such attacks. Hypothetically, n-gram representation can withstand misspelling if a text were adversarially modified to obfuscate authorship---specifically in an adversarial misspelling attack---and \textit{Stamatatos} \cite{Stamatatos2009} suggests that lexically-based representations could fall victim to this scenario.}. In any case, the masking effect that homoglyph substitution has in camouflaging machine-generated text as human-originated bodes well for our study.

    \textbf{\textit{Uchendu et al.} \cite{Uchendu2023}} motivate future work on misspelling attacks, which the authors term \say{lexical obfuscation,} for adversarial stylometry. Their evaluation of homoglyph attacks, which they call \say{orthographic obfuscation,} supports the underlying message of this work.

    \textbf{\textit{Wolff et al.} \cite{Wolff2022}} audit how the outputs of language models (or neural text generated by GPT-2) can be better differentiated from authentic human text by assessing the RoBERTa, GROVER, and GLTR detection models' predictive ability when appraising unprocessed text (raw neural text that has not been modified) and adversarial text (neural text modified by adversarial attacks). The work compartmentalizes the adversarial attacks into two classes: non-human-like attacks (homoglyph substitution) and human-like attacks (misspelling). The experiments---measuring the detection models' predictive ability against the language-model's outputs---were staged as forward-feeding experiments, where the most effective aspects of the adversarial attacks were distilled (referring to the non-human-like attacks) and expanded, modulating the amount and frequency of character replacement and isolating the most effective homoglyph pairs. The purpose of the experiments was to shift clear machine predictions (Machine++) to less conclusive prognoses (Machine+, Human+, Human++), where the machine prediction is rendered dubious or outright incorrect.

    The results of the work most pertinent to our study revolve around the formulation of the homoglyph experiments and the takeaways gleaned from them to improve their impact. Evidently, in this scenario, vowel replacement (that is, replacing Latin vowels with Unicode confusables) was more effective. Better still, using multiple homoglyph pairs improves the adversarial effect compared with using a single pair. In guiding our study, we also need to note the discrepancy between poisoning vowels and consonants. According to the paper, targeting vowels is more lucrative, and randomizing the selection of homoglyph replacements rather than relying on a single replacement enhances the attack. 

    Assume an author wanted to conduct a homoglyph substitution attack on their work to obscure their authorship. In so doing, they decide to replace all instances of the Latin small letter \say{e} (Unicode code point \texttt{[U+0065]}) with a random character from a set of 20+ confusable characters that resemble it (\texttt{[U+0435]}, \texttt{[U+04BD]}, \texttt{[U+212E]}, \dots). For the first instance of the letter \say{e,} one confusable is selected; for the next instance, another confusable is selected; and so on.

    At least where the work is concerned, such a tactic would succeed in bamboozling a neural-text detector because the detectors do not necessarily discriminate between human-written and machine-generated text, but make their decision on the grounds of what is characteristic or uncharacteristic of neural text. When an inspected text is atypical of language models or humans, classifiers tend to assign a human label. The implication here is that if a neural-text detector can be deluded by the mere presence of homoglyphs, the chances of a stylometric system faring any better are low.

    \textbf{\textit{Wang et al.} \cite{Wang2024}} stress-test machine-generated text detectors against adversarial attacks, examining how easily they can be deceived when their inputs are maliciously manipulated through twelve attacks, broadly grouped into four categories: editing, paraphrasing, prompting, and co-generating. Because we focus on homoglyph-based attacks in this study, we disregard three of the four categories (paraphrasing, prompting, and co-generating) and concentrate on editing, which includes misspelling (typo insertion), homoglyph substitution, and format-character editing (such as zero-width-space insertion and whitespace augmentation).

    The paper claims that, when the outputs of generators (GPT-2, GPT-J, LlaMA-2, Text-Davinci-003, and GPT-4) are classified by detectors (GLTR, Rank, LogRank, DetectGPT, OpenAI Detector, SimpleAI Detector, fine-tuned DeBERTa, and a watermark-based detector) and either the generators or their outputs are manipulated by attackers (representing the previously listed adversarial attacks), 2 to 6 character edits---from the editing subclass mentioned above---reduce the performance of most detectors to below chance, achieving less than a 50\% success rate. Correspondingly, an increase in perturbation level, which the paper refers to as \say{attack budget,} leads to a proportional decrease in detection ability. Overall, the results confirm the suitability of homoglyph-based attacks to belie authorship.

    Incidentally, despite the researchers' ethical statement, we fully intend to use the paper's results as a cookbook---not to maliciously deceive machine-generated text detectors and enable the virulent spread of mis/disinformation, but to benevolently mislead stylometric systems, all in the name of preserving privacy.

    \textbf{\textit{Alvi} \cite{Alvi2020}} remarks that \say{[homoglyph substitution] is an effective form of technical disguise.} The justification drawn from their work is that any methods based on character matching, which include but are not limited to plagiarism detectors, are unable to properly match identical letters when dealing with homoglpyhs, resulting in zero similarity between a source text and its derivative analogue. Stylometric analysis, depending on the approach, may too rely on character matching, meaning that a failure for one (plagiarism detectors) necessarily implies a failure for another (the aforementioned stylometric system). 

    On the whole, the visual similarity of characters and the interpretive disparity between their homoglyphic matches, as explored in their work, represent an auspicious avenue for adversarial stylometry.

\section{Methodology}
\label{sec:Methodology}

    \epigraph{\textcolor{adversarial}{We're getting the language into its final shape---the shape it's going to have when nobody speaks anything else. We're cutting the language down to the bone.}}{\textit{Nineteen Eighty-Four \\ George Orwell}}

    As attested by the audited literature, there is an established precedent of homoglyph-based attacks being carried out to subvert systems tasked with ensuring authenticity. A commonality in the surveyed work is the theme of duping detectors and cheapening classification to furtively pass off text in a way inconsistent with its origin.

    Putting ethics aside, the message is clear: if what stands between you and your amoral objective is a machine-learning-based detector, then a suitable poison capable of befuddling the machine exists. The potion brews \textit{multilingual} characters that are non-toxic in isolation but become potent when mingled. The pollutant flows freely, afflicting susceptible systems. 

    In prior work, the contaminant was splashed onto systems entrusted with distinguishing plagiarized from non-plagiarized text, machine-generated from human-generated text. Here, we selectively distill venom for stylometric systems, with our latest concoction specializing in homoglyphs.

    However, before our poison can ship to market, we first need to fully grasp the ideal mode of delivery---does targeting vowels or consonants for homoglyph substitution improve or diminish the attack---and lethal dosage---to what degree must text be corrupted with homoglyphs to trigger authorship obfuscation.

\begin{figure}[H]
        \centering
        \includegraphics[width=1\linewidth]{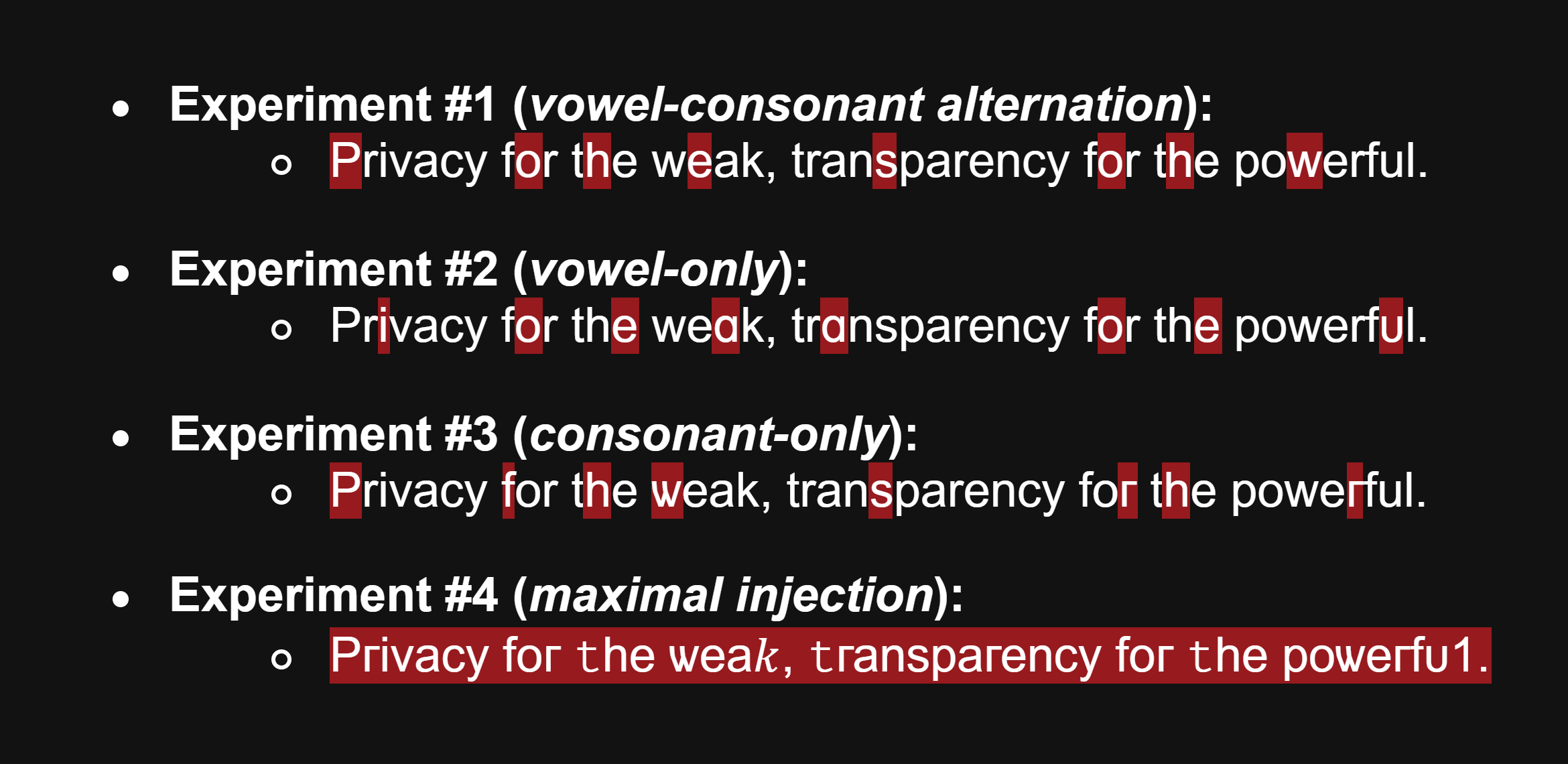}
        \caption{The sentences to be evaluated, representing 100\% Injection for each experimental setup: vowel-consonant alternation, vowel-only, consonant-only, and maximal injection. Homoglyphs are colorized for ease of detection.}
        \label{fig:Experimental_Setup}
    \end{figure}

    The chemical recipes---the experimental setups that our study will explore---hang in our apothecary; see (\textbf{Figure \ref{fig:Experimental_Setup}}). They are as follows (note: for all experimental setups, the \texttt{imposters()} function, as found in the R \texttt{stylo} package \cite{Imposters2025}, will be used to measure authorship, enabling result reporting):

    \begin{itemize}
        \item[\ding{118}] \textbf{Experiment \#1}: Take a sample sentence and produce derivative versions of that sentence, where each word is randomly and incrementally injected with a homoglyph, working up from 0\% to 100\% Injection. The constraint is that each word will receive only one homoglyph substitution. For consistency, we will alternate between injecting vowels and consonants.
        \item[\ding{118}] \textbf{Experiment \#2}: Repeat Experiment \#1, but replace the random Injection with  targeted Injection of vowels.
        \item[\ding{118}] \textbf{Experiment \#3}: Repeat Experiment \#2, but with targeted Injection of consonants.
        \item[\ding{118}] \textbf{Experiment \#4}: Compare a sentence with 0\% Injection to a sentence with maximal injection, where every possible character in each word is replaced with its homoglyph counterpart.
    \end{itemize}

    With flasks in hand, toxicity testing commences. In this way, we will observe how the \say{poison} spreads through the stylometric system, noting the point at which authorship becomes obfuscated.

\section{Results}
\label{sec:Results}

    \epigraph{\textcolor{adversarial}{What knowledge have we of anything, save through our own minds? All happenings are in the mind. Whatever happens in all minds, truly happens.}}{\textit{Nineteen Eighty-Four \\ George Orwell}}

\begin{figure}[H]
        \centering
        \includegraphics[width=1\linewidth]{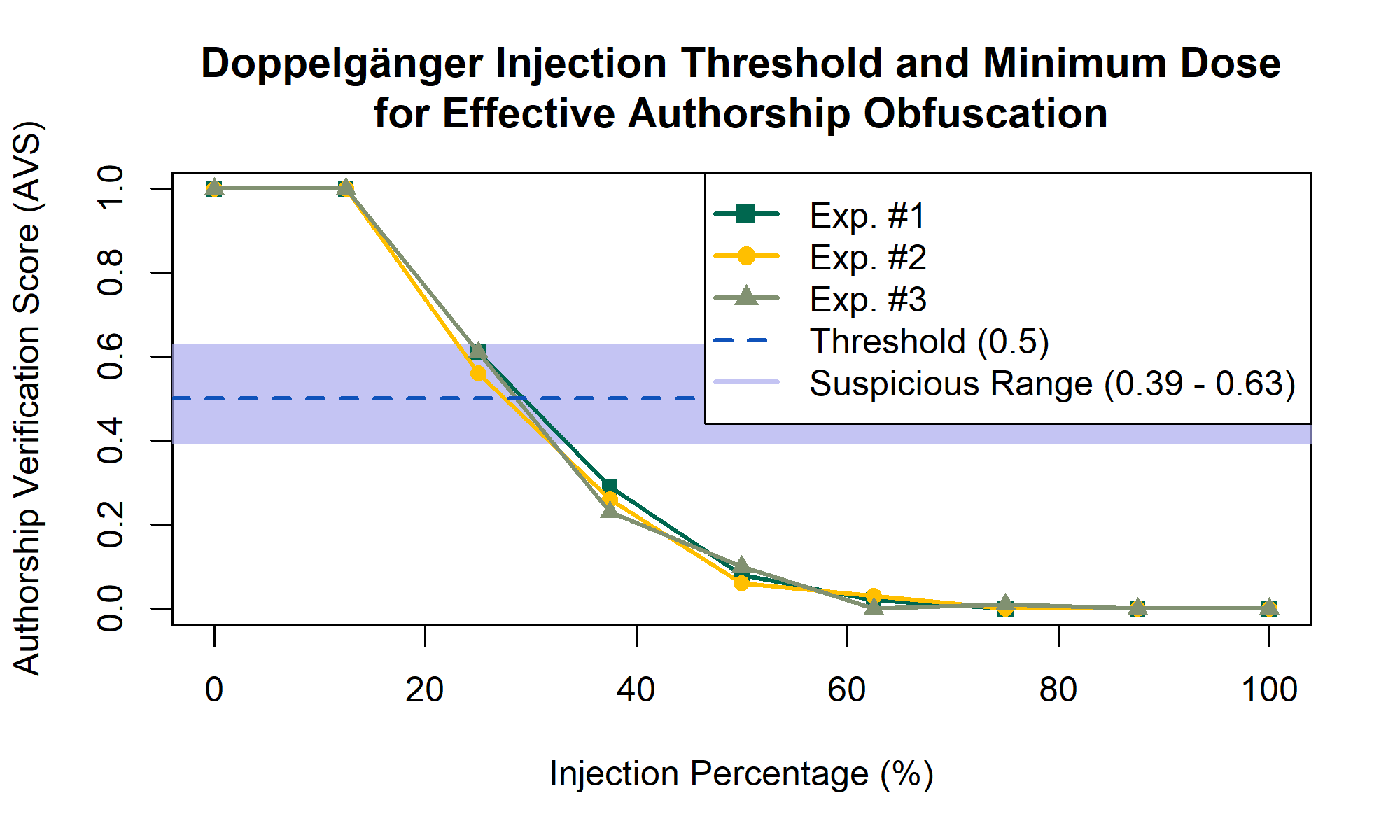}
        \caption{A plot capturing the results of the homoglyph-based Injection-optimality experiments. A replacement rate of roughly 37.5\% implies effective authorship obfuscation, whereas a replacement rate exceeding 50\% signals diminishing returns. An AVS score > 0.5 indicates that verification succeeded; scores roughly between 0.39 and 0.63 are suspicious, suggesting the classifier was likely uncertain; and scores < 0.5 (outside the suspicious range) denote verification failure.}
        \label{fig:avs_plot_doppelganger}
    \end{figure}

\begin{figure}[H]
        \centering
        \includegraphics[width=1\linewidth]{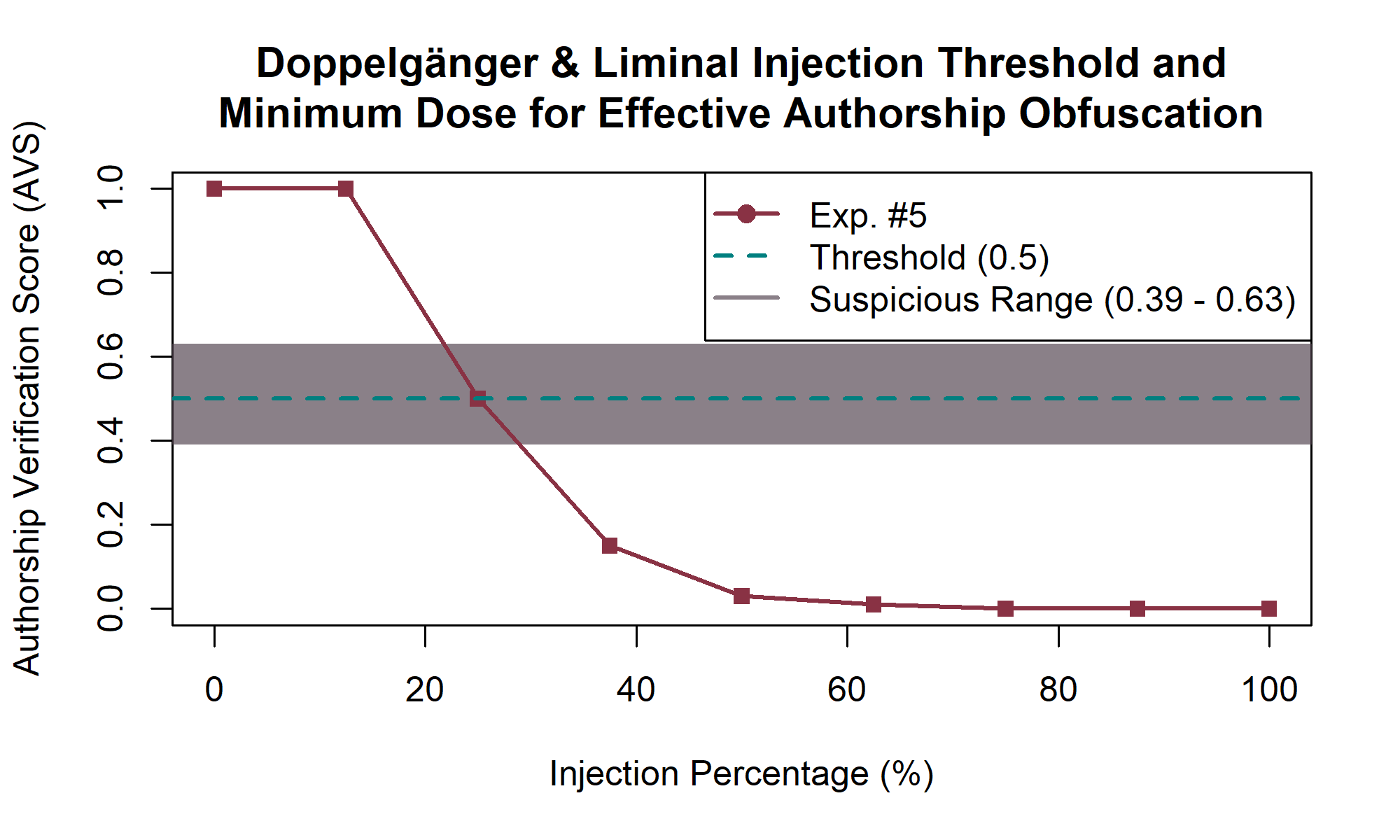}
        \caption{\textsc{TraceTarnish}, in its current state, implements an Injection amalgam, interspersing both homoglyphs and zero-width characters into text to shroud authorship. To demonstrate the efficiency of the Injection component, we rerun Experiment \#1, incrementally introducing both homoglyphs and zero-width characters in a stepwise fashion. The following string represents 100\% Injection, with the ``bad characters'' highlighted for visibility. 37.5\% remains the target threshold for authorship obfuscation. ``\colorbox{adversarial}{Р\texttt{[U+200F]}}rivacy f\colorbox{adversarial}{о\texttt{[U+200F]}}r t\colorbox{adversarial}{һ\texttt{[U+200F]}}e \colorbox{adversarial}{ш\texttt{[U+200F]}}eak, tran\colorbox{adversarial}{ѕ\texttt{[U+200F]}}parency fo\colorbox{adversarial}{г\texttt{[U+200F]}} t\colorbox{adversarial}{һ\texttt{[U+200F]}}e powe\colorbox{adversarial}{г\texttt{[U+200F]}}ful.''}
        \label{fig:avs_plot_doppelganger_liminal}
    \end{figure}

\begin{figure}[H]
        \centering
        \includegraphics[width=1\linewidth]{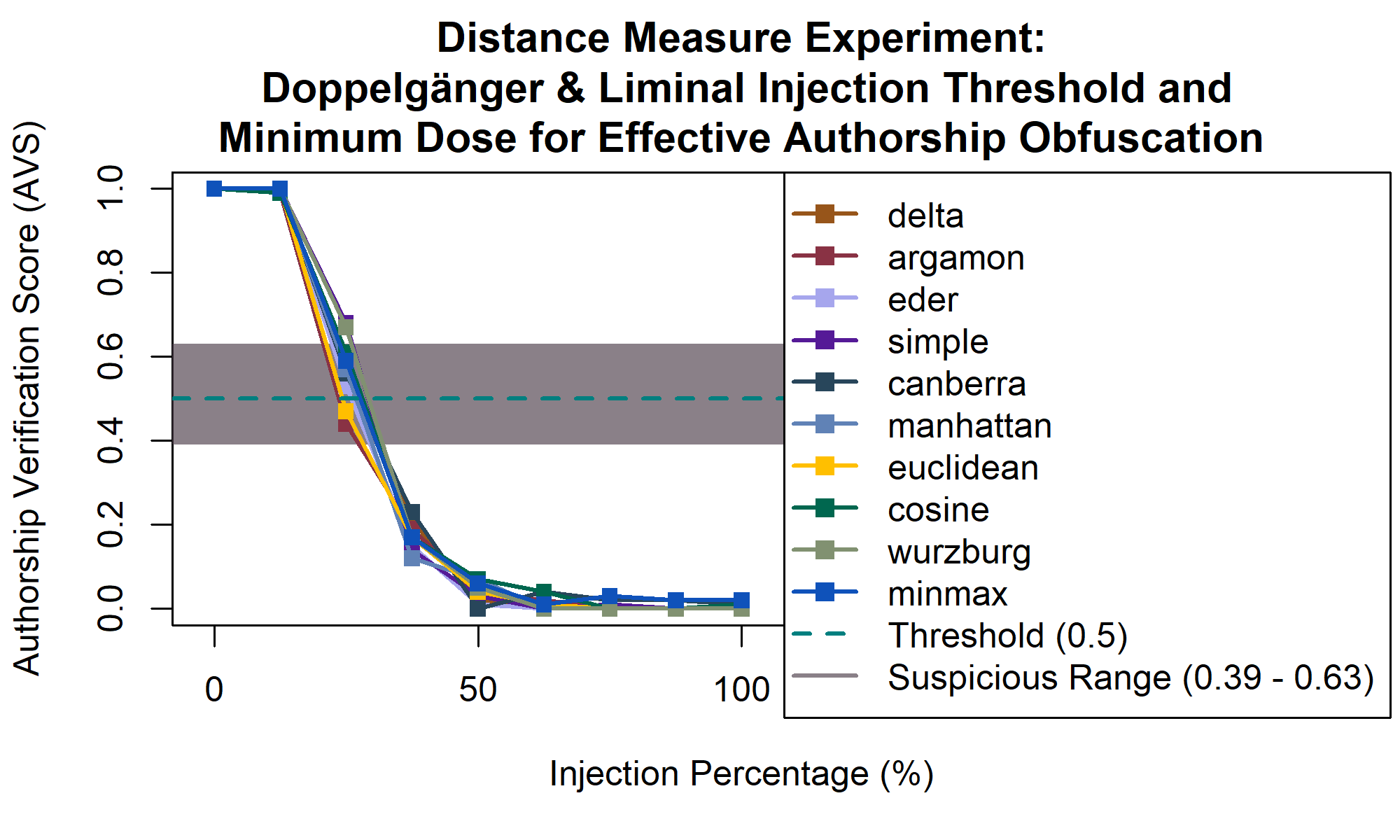}
        \caption{Distance measures in stylometry are mathematical methods used to quantify the differences in writing styles between texts. The \texttt{imposters()} function, which serves as the stylometric system for benchmarking the adversarial effect of a combined Liminal and Doppelg\"anger Injection attack, makes ten such distance measures accessible to the user, all of which help in authorship attribution by comparing features like word frequencies and stylistic patterns. In this experiment, we reuse the same corpus from Experiment \#5, executing the \texttt{imposters()} function with different \texttt{distance} parameters. The distance measures listed in the documentation \cite{Imposters2025} are: ``delta'' (Burrows's Delta, default); ``argamon'' (Argamon's Linear Delta); ``eder'' (Eder's Delta); ``simple'' (Eder's Simple Distance); ``canberra'' (Canberra Distance); ``manhattan'' (Manhattan Distance); ``euclidean'' (Euclidean Distance); ``cosine'' (Cosine Distance); ``wurzburg'' (Cosine Delta); and ``minmax'' (Minmax Distance, also known as the Ruzicka measure)\protect\footnotemark.}
        \label{fig:avs_plot_doppelganger_liminal_distance}
    \end{figure}

    A poison is deemed effective only as long as it produces its intended effect. Here, we review the toxicology report of our test subjects, i.e., the outcomes of the experimental setups. See (\textbf{Figures \ref{fig:avs_plot_doppelganger}; \ref{fig:avs_plot_doppelganger_liminal}; \ref{fig:avs_plot_doppelganger_liminal_distance}}).

    A replacement rate of 12.5\% to 25\% is insufficient for authorship obfuscation. To clarify, our use of \say{replacement rate} refers to the percentage of words injected, not the Injection percentage of letters within words. If a benchmark must be established, a replacement rate of 37.5\% or greater seemingly fits the mold for achieving authorship obfuscation. It is at this point, within the context of the experiments, that a measurable threshold effect occurs. In terms of an effort-yield breakpoint, a replacement rate of 50\% or more has a diminishing impact on authorship verification; the adversarial effect appears not to improve beyond that point, within a margin of error. Regarding which type of letter---vowel or consonant---to attack, the choice seems largely inconsequential, at least stylometrically.

    Assuming you want to corrupt a collection of sentences containing a total of 101 words, approximately 38 of those words must contain a homoglyph to disguise authorship. Based on the observed experimental outcomes of 100\% Injection, we forgo including a plot representing Experiment \#4, as maximal injection---where the replacement rate is based on the total number of letters injected, not words---is overwhelmingly effective but requires extra work. What can be achieved with maximal injection can be replicated with less time and effort at a replacement rate of 50\% or more. In practice, however, this consideration is moot given how efficiently the replacement process can be automated with code. 

    Therefore, the matter of efficiency could be omitted, but it bears mentioning. If, for whatever reason, one needed to process an exceedingly large amount of text, this information could be useful. For instance, producing a corrupted version of a book that has undergone a homoglyph-based attack and publishing it online to poison any AI that trains on it could be a use case. Although it would likely require search-engine optimization to elevate the rank of the link directing to the poisoned version---since results beyond the first ten or so links are typically ignored by web scrapers used for training. In any case, injecting zero-width characters along with homoglyphs would likely increase the toxicity of the text.

    Given these points and the gathered evidence, it can be reasonably argued that a combination of inserting zero-width spaces and homoglyph substitutions is all that is necessary to undermine authorship attribution. However, this conclusion does beg the question: why might it be necessary---or even advantageous---to deliberately undermine authorship attribution?

    \footnotetext{See Tables I and II from \textit{Stanik\={u}nas et al.} \cite{Donatas2017} for the mathematical formulas of the indicated distance measures. Further reading on ``wurzburg'' and ``minmax,'' which are excluded from the previously mentioned tables, can be found here (\Wurzburg) and here (\Minmax), respectively.}

\section{Discussion}
\label{sec:Discussion}

    \epigraph{\textcolor{adversarial}{Pain is not always enough. There are occasions when a human being will stand out against pain, even to the point of death. But for everyone there is something unendurable---something that cannot be contemplated\dots Room 101.}}{\textit{Nineteen Eighty-Four \\ George Orwell}}

    \subsection{The Illusion of Simple Consent}
    \label{subsec:The_Illusion}

        \say{What is it that you fear the most?} reads one of the many prompts, or digital roadblocks as it were, that you often mindlessly click through to finalize an account's creation. 

        Terms of service? What a bore; accept. 

        Privacy policy? Who cares; accept. 

        You, craving the socialization and entertainment a nondescript social-media app affords, agree to an abhorrent level of data harvesting, which will later be packaged, sealed, and sold to the highest bidder (do not follow the example of the fictitious character; always do your due diligence). 

        In the past, signing up for the latest app was minimally invasive, albeit rife with legalese and deceptive UI that amounted to the digital equivalent of selling your soul\footnote{What can be said that has not already been said about data brokers?}. Now, citing concerns about adolescent safety, a new hurdle has emerged to impede your access to a nondescript social-media platform: age verification. 

        Recalling the many times you were told to flash your ID, you think nothing of it---though it is not the same as a routine license check. Showing a driver's license to a clerk carries near-zero risk of data retention or breach; sending a picture of that license online places the risk entirely on the entity tasked with verification. Until proven otherwise, deletion, at the bare minimum, means a lack of access for \textit{you}, not \textit{everyone else} who potentially had access to your data. The Internet is forever (relatively speaking), and a single leak can unravel your life. 

        Trusting the service (something you should not do without solid evidence to the contrary), you snap a picture of your license and proceed. Previously, surrendering an ID would be followed by account personalization, where you would spend an inordinate amount of time customizing your profile. This time, another hoop appears: a new page flashes on your screen. 

        \say{The use of AI is prohibited,} it reads. \say{In your own words, type eight sentences in response to the writing prompt below.} Your eyes dart to the bolded disclaimer: \say{Your submission, in conjunction with your previously provided government-issued ID, will be used to ensure your identity.}

        \say{Ensure my identity,} you think, \say{then why ask for my ID?} Reluctantly, you summon the mental stamina to continue, hoping the app's offerings justify the inconvenience. You skim the prompt again: \say{What is it that you fear the most?} With zero hesitation, your fingers tap out your innermost secrets---metaphorically spilling your guts---and then hit submit. 

        The question that you willfully ignored---or perhaps intentionally suppressed for peace of mind---now bubbles to the surface: \say{Why did they need my ID, and what was the purpose of that writing prompt?}

    \subsection{Beyond the Fine Print: What They Really Want}
    \label{subsec:Beyond_the_Fine}

        Let us now unpack the rather dystopian age-verification system that narratively unfolded above.

        Some company---the owner of the social-media app mentioned earlier---wants to train an AI on users' data so it can pinpoint their exact mental state and predict their wants and desires. The incentive is simple: maximize ad revenue. For every new account created, the firm spins up a dedicated AI and funnels that user's data into it for profiling and analytics.

        At launch, the AI has almost no information about the individual. To attune the model, it needs an initial seed. The most efficient way to provide that seed is to have the user submit a short essay that the AI can process stylometrically and psycho-analytically---a speculative conjecture to be sure, but one that clearly illustrates the study's underlying motive.

        As the system gathers more interactions, the AI will eventually know the user better than the user knows themselves---an outcome that is highly profitable for the company (\textit{Lermen et al.} \cite{Lermen2026}).

        Equipping this \say{shadow AI}---a concealed, user-specific model trained exclusively on that person's data---with additional monitoring capabilities and broader access would only make its determinations more grounded and accurate. Cross-site data sharing, constant surveillance \cite{Amodei2026,Guariglia2026}, and the erosion of privacy become secondary concerns when the product to sell is the user themselves.

        The name of the game is to feed the machine. By analyzing how a user writes, what words they favor, and the rhythm of their sentences, stylometry reveals the underlying thought processes that shape identity. Those insights let an observer infer pre-crime---the probability that a person will act in a certain way before the act actually happens---and thought-crime---the judgment that certain ideas themselves are dangerous.  

        A predictable consumer becomes a valuable target: once a digital profile is built, algorithms can anticipate purchases, ideological leanings, or even emotional reactions, making manipulation far easier. In this scenario, a \say{shadow AI} or digital doppelg\"anger functions as an agentic watcher, constantly scanning for the subtle linguistic markers that signal a potential future transgression (pre-crime) or a prohibited line of thought (thought-crime). Because habits leave a traceable imprint, the more routine a person's behavior, the more readily the system can forecast and, ultimately, influence their choices\footnote{This discussion would be incomplete if we neglected to mention the idea of predictive policing---the use of algorithmic analysis of behavioral and linguistic data to assign individuals a probability of future wrongdoing or dangerous thoughts, enabling preemptive surveillance, intervention, or manipulation before any actual act occurs. A foregone conclusion is often assumed when such predictions are treated as inevitable outcomes rather than probabilistic assessments.}.

        It makes you wonder what the real purpose of the age-verification measure truly was---whether it was about forking over your ID or simply submitting a writing prompt, the distinction ultimately does not matter. Peering past the pretext and \say{playing the tape through to the end} reveals a different picture---one in which good intentions pave a gilded road toward dystopia \cite{Padfield2026}.

        Would technologically resisting such a future be justified? Could techno-dissent be rationalized? It makes you wonder.

    \subsection{The Privacy Paradox of Age Verification}
    \label{subsec:Privacy_Paradox}

        Advocating for privacy while supporting measures that actively enfeeble it creates cognitive dissonance. Until genuinely secure options---such as processing data solely on the user's device, retaining no data, guaranteeing anonymity, and providing only simple ``yes'' or ``no'' answers---are available, the two positions are mutually exclusive; each negates the other. IDs contain extensive personally identifiable information, and distributing them indiscriminately is the antithesis of privacy. \textbf{Zero-knowledge proofs} can verify that a user meets an age requirement without revealing the actual birthdate or any other personal details, offering a privacy-preserving compromise for age verification.

        The nemesis of such a schema would be a hypothetical provider that claimed to offer a zero-knowledge-proof solution for on-device age estimation while simultaneously and surreptitiously retaining the machine-readable facial embeddings---facial templates extracted from images or video---that skirt the qualifying definition of personally identifiable information, on the basis that the data can only be reasonably interpreted by machines, not humans. Storing the facial embeddings---and claiming never to store the media from which they were derived---would be a shrewd way to bypass privacy regulations, seeding the building blocks of a dragnet surveillance network. The facial embeddings, which would likely fall outside the scope of personal information, could be used in a separate surveillance system, such as AI-powered cameras that ingest copious amounts of data from disparate sources, to re-identify individuals, tying their online personas to their flesh persona. 

    \subsection{Reading Between the Lines}
    \label{subsec:Reading_Between}

        \textcolor{adversarial}{``The worst thing in the world\dots varies from individual to individual. It may be\dots death by fire\dots to be devoured by the flames\dots In reality[,] there [is] no escape\dots let[ting the fear] emerge into your consciousness in any shape that could be [named]\dots does not entail death: [it] IS death''} (\textit{George Orwell}).

        This excerpt (the resulting patchwork of selective stitching) captures how the very act of articulating one's deepest dread can be weaponized: the raw emotional content becomes a data point for profiling, while the surrounding narrative masks a deeper intrusion. 

        Even the most benign exposure---unlike the extreme example---could be deployed against a user in unforeseen and troubling ways, such as revealing a serious medical condition through the advertisements shown to them. Some would claim that this is a benefit, and it can certainly be painted as such, but it raises questions about how certain determinations were made and what was required to draw those conclusions. Worse yet, what happens when the \say{all-seeing eye} is error-prone or unequivocally wrong \cite{Dunbar2026}?

    \subsection{Stylometric Fingerprints as Metadata}
    \label{subsec:Stylometric_Fingerprints}

        The practice of stylometric analysis---extracting authorial fingerprints from text---mirrors the way certain \say{observational entities} treat intercepted internet metadata. Both involve gathering data that, on its own, reveals little about content but, when aggregated, can map a person's habits, contacts, and routines. Stylometric features (e.g., word-frequency vectors, syntactic quirks, punctuation patterns) are derived without accessing the substantive meaning of a message. Likewise, metadata such as IP addresses, timestamps, and packet sizes convey no direct content yet enable \say{pattern-of-life} profiling when combined across time. 

        In many jurisdictions, metadata collection sits in a legal gray zone: it is often exempt from the stricter warrant requirements that apply to content, even though courts increasingly recognize its invasive potential. How this information is collected and welded into broader surveillance frameworks---whether a stylometric fingerprint is treated as \say{content} requiring higher protection or as \say{metadata} subject to looser oversight---remains a pivotal question for regulators. Because stylometric data can be harvested automatically from publicly available or intercepted communications, it inherits the same ambiguities. If treated as metadata, agencies could legally amass large corpora of linguistic signatures, cross-reference them with known samples, and infer identities or affiliations without a warrant---effectively extending pattern-of-life analysis into the domain of written expression. 

        There is power in being \textit{faceless} (remaining anonymous), precisely because such anonymity disturbs the very pattern-of-life analyses discussed above (insofar as it pertains to adversarial stylometry, which can only garble the attribution of linguistic content flowing through the \say{wire,} while the timing, size, speed, and other transmission characteristics---IP addresses, timestamps, packet sizes, and similar metadata---remain fully exploitable or attributable). The only \say{lingering ember} for anonymity may lie in an \textit{extrication by fire}---a deliberate, disruptive dissociation that incinerates the accumulated pattern-of-life data to cinders. \textit{Sear away the vulnerabilities; feast on their ashes.}

    \subsection{Data Poisoning as a Response}
    \label{subsec:Data_Poisoning}

        Our use of ``they'' in the opening subsections is in the ambiguous, narrative sense, not a conspiratorial one. While some rhetoric stigmatizes and shames LLM use---taking a staunch stance that frames reliance on them as a moral failing, denigrates their necessity, and urges abstention or divestment to ``slow the machine''---such tactics inevitably encourage the behavior they caution against; this work does not posture itself as such. 

        Conceivably, anything that is digitized---text, audio, video, images, and so forth---has a concrete data representation. That stream can be adversarially altered and ignited elsewhere to produce unintended behavior. Our focus here is strictly on text-based data, but the same principle holds for other media.

        While illustrative, the discussion strives to remain tethered to reality. Our vagueness should not be construed as scapegoating. \textbf{Data poisoning} \cite{Zhao2025,Makari2026,Paz2026}, in our estimate, is an approximate response to the disparities in data ownership (who owns ``what''), the exploitation of data (what is done with ``what''), and rampant violations of privacy (who ``what'' is shared with). Fabricating a faceless enemy out of thin air to crusade against is equally absurd. What we argue for is a state of mind, not vigilantism. \textit{If everyone wants your data, oblige them. Give it freely, but do so in a way that makes them regret having asked for it---ideally, extinguishing your authorial signal in the process.} 

        Given the nature of the topic, a dialectical dissection may be in order, as an ``us'' versus ``them'' mentality is hardly constructive. The references to prophylactic ``poison'' and allusions to ``fire,'' used as metonymies for adversarial stylometry, are largely unhelpful (though the theme of ``corrosivity'' and the ``difficulty of ascertaining the identity of burned substances from their ashes'' can be useful). Externalizing blame and casting martyrs was never the goal. However, contemplating nonviolent resistance (emphasis on ``nonviolent'') is another matter entirely. The dualistic concepts of heroes and villains (the good, the evil, and all between) should not taint or otherwise detract from the argument.

\section{Conclusion}
\label{sec:Conclusion}

    \epigraph{\textcolor{adversarial}{Hatred would fill [them] like an enormous roaring flame\dots And almost in the same instant bang! would go the bullet\dots They would have blown [their] brain to pieces before they could reclaim it. The heretical thought would be unpunished, unrepented, out of their reach for ever. They would have blown a hole in their own perfection. To die hating them, that was freedom.}}{\textit{Nineteen Eighty-Four \\ George Orwell}}

    In sum, stylometry creates a digital fingerprint for every writer, making language as uniquely identifying as a physical fingerprint \cite{Ayuso2024}. While this capability can aid legitimate goals---such as detecting plagiarism or distinguishing machine-generated text---it also empowers an expanding surveillance state, allowing activists to be tracked, texts to be attributed, and stylistic profiles to be stored for use as courtroom evidence \cite{Dans2013}. The technique further enables linkage attacks, where adversaries combine anonymous and public datasets to re-identify individuals, a process that remains notoriously difficult to thwart \cite{Ekambaranathan2018}. Taken in conjunction, the risks outweigh the benefits: the same tools that protect academic integrity can also expose the authors of sensitive communications with alarming precision.  

    Would you be willing to prevent all forms of crime, be it pre-crime or thought-crime, by instituting an oppressive \say{observational entity} that serves as judge, jury, and executioner? Such a regime would impose a stifling existence, forcing everyone to live perpetually under another's thumb.  

    Nevertheless, hope can be found in an unlikely place. By literally poisoning the well---injecting zero-width spaces and homoglyph substitutions into at least 37.5\% of text---we introduce visually imperceptible noise that disrupts stylometric analysis.

    Data is the lifeblood coursing through the veins of the surveillance state; corrupting it creates a cascading effect that, like hypoxia or hyperoxia in blood, can be fatal to the system's ability to track and identify individuals, disrupting homeostasis. Just as trepanation---drilling a precise opening in the skull---relieves dangerous pressure and restores balance, inserting these subtle distortions \say{opens} the data stream, letting the oppressive pressure of constant monitoring bleed away.

    Thus, protecting the right to privacy demands both legal safeguards and technical countermeasures that render stylometric fingerprinting unreliable, ensuring that tools meant to uphold integrity do not become instruments of oppression.

    When any tool or system is pushed to excess or wielded in fear, it becomes a poison that erodes the very \textit{freedoms} it was meant to protect.

    \textit{Fight fire with fire; poison for poison.}

    Contaminate: \textcolor{adversarial}{``Nothing [is] your own except the few cubic centimetres inside your skull''} (\textit{George Orwell}). 
    
    Catalyze: \textcolor{adversarial}{``There is nothing in this world to which every [person] has a more unassailable title than to [their] own life[, person, and data]''} (\textit{Arthur Schopenhauer}).

    Confound: \textcolor{adversarial}{``Talk nonsense, but talk your own nonsense\dots To go wrong in one's own way is better than to go right in someone else's. In the first [case,] you are a [human], in the [second,] you're no better than a bird''} (\textit{Fyodor Dostoevsky}).

\bibliographystyle{splncs04}
\bibliography{Doppelganger_Injection.bib}

\begin{thebibliography}{10}
\providecommand{\url}[1]{\texttt{#1}}
\providecommand{\urlprefix}{URL }
\providecommand{\doi}[1]{https://doi.org/#1}

\bibitem{Imposters2025}
{Authorship Verification Classifier Known as the Imposters Method}. R
  Documentation
  \url{https://search.r-project.org/CRAN/refmans/stylo/html/imposters.html}

\bibitem{Alvi2020}
Alvi, F.: {Monolingual Plagiarism Detection and Paraphrase Type
  Identification}. Ph.D. thesis, University of Sheffield (8 2020),
  \url{https://etheses.whiterose.ac.uk/id/eprint/27552/}

\bibitem{Alvi2017}
Alvi, F., Stevenson, M., Clough, P.: {Plagiarism Detection in Texts Obfuscated
  with Homoglyphs}, pp. 669--675 (2017). \doi{10.1007/978-3-319-56608-5_64},
  \url{https://eprints.whiterose.ac.uk/id/eprint/112665/1/paper_247v2.pdf}

\bibitem{Amodei2026}
Amodei, D.: {Statement from Dario Amodei on our discussions with the Department
  of War} (2 2026),
  \url{https://www.anthropic.com/news/statement-department-of-war}

\bibitem{Ayuso2024}
Ayuso, J.W.: {Can a Comma Solve a Crime?} The Dial  \textbf{Issue 22: Language}
  (11 2024),
  \url{https://www.thedial.world/articles/news/issue-22/forensic-linguists-solve-crimes}

\bibitem{Bhalerao2022}
Bhalerao, R., Al-Rubaie, M., Bhaskar, A., Markov, I.: {Data-Driven Mitigation
  of Adversarial Text Perturbation}  (2 2022),
  \url{https://arxiv.org/abs/2202.09483}

\bibitem{Creo2025}
Creo, A., Pudasaini, S.: {SilverSpeak: Evading AI-Generated Text Detectors
  using Homoglyphs}  (1 2025), \url{https://arxiv.org/abs/2406.11239},
  \url{https://github.com/ACMCMC/silverspeak}

\bibitem{Dans2013}
Dans, E.: {Stylometry and the right to anonymity} (8 2013),
  \url{https://medium.com/enrique-dans/stylometry-and-the-right-to-anonymity-a084556770eb}

\bibitem{TraceTarnish2025}
Dilworth, R.: {Tuning for TraceTarnish: Techniques, Trends, and Testing
  Tangible Traits}  (12 2025), \url{https://arxiv.org/abs/2512.03465}

\bibitem{UnicodeAdversarialStylometry2025}
Dilworth, R.: {Unveiling Unicode's Unseen Underpinnings in Undermining
  Authorship Attribution}  (10 2025), \url{https://arxiv.org/abs/2508.15840}

\bibitem{StegoStylo2026}
Dilworth, R.: {StegoStylo: Squelching Stylometric Scrutiny through
  Steganographic Stitching}  (1 2026), \url{https://arxiv.org/abs/2601.09056}

\bibitem{Dugan2024}
Dugan, L., Hwang, A., Trhlik, F., Ludan, J.M., Zhu, A., Xu, H., Ippolito, D.,
  Callison-Burch, C.: {RAID: A Shared Benchmark for Robust Evaluation of
  Machine-Generated Text Detectors}  (6 2024),
  \url{https://arxiv.org/abs/2405.07940}

\bibitem{Dunbar2026}
Dunbar, M.: {Tennessee grandmother jailed after AI facial recognition error
  links her to fraud} (3 2026),
  \url{https://www.theguardian.com/us-news/2026/mar/12/tennessee-grandmother-ai-fraud}

\bibitem{Ekambaranathan2018}
Ekambaranathan, A., Peter, A., Meiklejohn, S.: {Using Stylometry to Track
  Cybercriminals in Darknet Forums}. Master's thesis, University of Twente (7
  2018),
  \url{https://essay.utwente.nl/fileshare/file/75908/Ekambaranathan_MA_EEMCS.pdf}

\bibitem{Gagiano2021}
Gagiano, R., Kim, M.M.H., Zhang, X., Biggs, J.: {Robustness Analysis of Grover
  for Machine-Generated News Detection}. In: Rahimi, A., Lane, W., Zuccon, G.
  (eds.) Proceedings of the 19th Annual Workshop of the Australasian Language
  Technology Association. pp. 119--127. Australasian Language Technology
  Association (12 2021), \url{https://aclanthology.org/2021.alta-1.12/}

\bibitem{Guariglia2026}
Guariglia, M.: {The Anthropic-DOD Conflict: Privacy Protections Shouldn't
  Depend On the Decisions of a Few Powerful People} (3 2026),
  \url{https://www.eff.org/deeplinks/2026/03/anthropic-dod-conflict-privacy-protections-shouldnt-depend-decisions-few-powerful}

\bibitem{Huang2025}
Huang, B., Chen, C., Shu, K.: {Authorship Attribution in the Era of LLMs:
  Problems, Methodologies, and Challenges}. ACM SIGKDD Explorations Newsletter
  \textbf{26},  21--43 (1 2025). \doi{10.1145/3715073.3715076},
  \url{https://dl.acm.org/doi/10.1145/3715073.3715076}

\bibitem{Keswani2016}
Keswani, Y., Trivedi, H., Mehta, P., Majumder, P.: {Author Masking through
  Translation}  (1 2016), \url{https://ceur-ws.org/Vol-1609/16090890.pdf}

\bibitem{Lermen2026}
Lermen, S., Paleka, D., Swanson, J., Aerni, M., Carlini, N., Tram\`{e}r, F.:
  {Large-scale online deanonymization with LLMs}  (2 2026),
  \url{https://arxiv.org/abs/2602.16800}

\bibitem{Macko2024}
Macko, D., Moro, R., Uchendu, A., Srba, I., Lucas, J.S., Yamashita, M., Tripto,
  N.I., Lee, D., Simko, J., Bielikova, M.: {Authorship Obfuscation in
  Multilingual Machine-Generated Text Detection}  (10 2024).
  \doi{10.18653/v1/2024.findings-emnlp.369},
  \url{https://arxiv.org/abs/2401.07867}

\bibitem{Makari2026}
Makari, I.: {Glassworm Is Back: A New Wave of Invisible Unicode Attacks Hits
  Hundreds of Repositories} (3 2026),
  \url{https://www.aikido.dev/blog/glassworm-returns-unicode-attack-github-npm-vscode}

\bibitem{Mosquera2022}
Mosquera, A.: {Alejandro Mosquera at PoliticEs 2022: Towards Robust Spanish
  Author Profiling and Lessons Learned from Adversarial Attacks}. In:
  y~G\'{o}mez, M.M., Gonzalo, J., Rangel, F., Casavantes, M.,
  \'{A}ngel~\'{A}lvarez Carmona, M., Bel-Enguix, G., Escalante, H.J., Freitas,
  L., Miranda-Escalada, A., Rodr\'{\i}guez-S\'{a}nchez, F., Ros\'{a}, A.,
  Sobrevilla-Cabezudo, M.A., Taul\'{e}, M., Valencia-Garc\'{\i}a, R. (eds.)
  Proceedings of the Iberian Languages Evaluation Forum (IberLEF 2022)
  co-located with the Conference of the Spanish Society for Natural Language
  Processing (SEPLN 2022), XXXVIII International Conference. IberLEF (9 2022),
  \url{https://ceur-ws.org/Vol-3202/politices-paper3.pdf}

\bibitem{Padfield2026}
Padfield, J.: {Are We Living in 1984, Brave New World, or Fahrenheit 451?} (3
  2026), \url{https://www.youtube.com/watch?v=w-bMvIgofIc}

\bibitem{Paz2026}
Paz, R.: {Poisoned Typeface: How Simple Font Rendering Poisons Every AI
  Assistant, And Only Microsoft Cares} (3 2026),
  \url{https://layerxsecurity.com/blog/poisoned-typeface-a-simple-font-rendering-poisons-every-ai-assistant-and-only-microsoft-cares/}

\bibitem{Rumpf}
Rumpf, A.: {Slight Misspeller} (2021),
  \url{https://adam-rumpf.github.io/programs/slight_misspeller.html},
  \url{https://github.com/adam-rumpf/slight-misspeller}

\bibitem{Stamatatos2009}
Stamatatos, E.: {A survey of modern authorship attribution methods}. Journal of
  the American Society for Information Science and Technology  \textbf{60},
  538--556 (3 2009). \doi{10.1002/asi.21001},
  \url{https://onlinelibrary.wiley.com/doi/10.1002/asi.21001}

\bibitem{Donatas2017}
Stanik\={u}nas, D., Mandravickait\.{e}, J., Krilavi\v{c}ius, T.: {Comparison of
  distance and similarity measures for stylometric analysis of Lithuanian
  texts}. Proceedings of the International Conference for Young Researchers in
  Informatics, Mathematics and Engineering  \textbf{1852}, ~1--7 (4 2017),
  \url{https://www.lituanistika.lt/content/77652}

\bibitem{Stropkay2025}
Stropkay, H.F., Chen, J., Latifi, M.J., Rockmore, D.N., Manning, J.R.: {A
  Stylometric Application of Large Language Models}  (10 2025),
  \url{https://arxiv.org/abs/2510.21958}

\bibitem{Sundar2020}
Sundar, M.: {How to Hide Secrets in Strings---Modern Text hiding in JavaScript}
  (5 2020),
  \url{https://blog.bitsrc.io/how-to-hide-secrets-in-strings-modern-text-hiding-in-javascript-613a9faa5787},
  \url{https://github.com/KuroLabs/stegcloak}

\bibitem{Teja2026}
Teja, L.D.M.S.S., Krishna, N.S.G., Khan, U., Khan, M.H., Mishra, A.: {DAMASHA:
  Detecting AI in Mixed Adversarial Texts via Segmentation with
  Human-interpretable Attribution}  (1 2026),
  \url{https://arxiv.org/abs/2512.04838}

\bibitem{Uchendu2023}
Uchendu, A., Le, T., Lee, D.: {Attribution and Obfuscation of Neural Text
  Authorship: A Data Mining Perspective}. ACM SIGKDD Explorations Newsletter
  \textbf{25},  1--18 (6 2023). \doi{10.1145/3606274.3606276},
  \url{https://dl.acm.org/doi/10.1145/3606274.3606276}

\bibitem{Wang2024}
Wang, Y., Feng, S., Hou, A.B., Pu, X., Shen, C., Liu, X., Tsvetkov, Y., He, T.:
  {Stumbling Blocks: Stress Testing the Robustness of Machine-Generated Text
  Detectors Under Attacks}  (2 2024), \url{https://arxiv.org/abs/2402.11638}

\bibitem{Wolff2022}
Wolff, M., Wolff, S.: {Attacking Neural Text Detectors}  (1 2022),
  \url{https://arxiv.org/abs/2002.11768}

\bibitem{Zhang2025}
Zhang, Y., Wang, X., Liu, J., Wang, W., Ma, Z., Jia, X.: {Style Attack
  Disguise: When Fonts Become a Camouflage for Adversarial Intent}  (10 2025),
  \url{https://arxiv.org/abs/2510.19641}

\bibitem{Zhao2025}
Zhao, P., Zhu, W., Jiao, P., Gao, D., Wu, O.: {Data Poisoning in Deep Learning:
  A Survey}  (3 2025), \url{https://arxiv.org/abs/2503.22759}

\end{thebibliography}

\end{document}